\documentstyle[aps
	,twocolumn
	]{revtex}
\begin{document}

\onecolumn

\title{\bf Models for diffusion and island 
growth in metal monolayers}
\author{Ofer Biham\footnote{Tel:972-2-6584363;Fax:972-2-6520089;Email:biham@flounder.fiz.huji.ac.il}
and Itay Furman}
\address{
Racah Institute of Physics,
The Hebrew University,
Jerusalem 91904,
Israel}
\author{Majid Karimi}
\address{
Physics Department,
Indiana University of Pennsylvania,
Indiana, PA 15705
}
\author{Gianfranco Vidali, Rosemary Kennett and Hong Zeng}
\address{
Department of Physics,
Syracuse University,
Syracuse, NY 13244
}
\maketitle

\begin{abstract}
A model that describes self diffusion, island
nucleation and film growth on FCC(001) metal substrates
is presented.
The parameters of the model are optimized to describe
Cu diffusion on Cu(001), by comparing activation energy barriers
to  a full set of barriers obtained 
from semi-empirical potentials
via the embedded atom method.  
It is found that this model (model I), 
with only three parameters,
provides a very good description of the full landscape
of hopping energy barriers. 
These energy barriers are grouped in 
four main peaks.
A reduced model (model II) with only two parameters,
is also presented, in which 
each peak is collapsed into a single energy value.
From the results of our simulations, 
we find that this model
still maintains the essential features of
diffusion and growth on this model surface.
We find that hopping rates along
island edges are much higher than for isolated atoms
(giving rise to compact island shapes) and 
that vacancy mobility is higher
than adatom mobility.
We observe substantial dimer mobility 
(comparable to the single atom mobility) 
as well as some mobility of trimers.
Mobility of small islands
affects the scaling of
island density $N$ vs. deposition rate $F$, 
$N \sim F^{\gamma}$,
as well as the island size distribution. 
In the asymptotic limit
of slow deposition, scaling arguments and rate equations
show that
$\gamma=i^{\ast}/(2 i^{\ast} + 1)$
where $i^{\ast}$ is the size of
the largest mobile island. 
Our Monte Carlo results, obtained for a 
range of experimentally relevant conditions, show
$\gamma = 0.32 \pm 0.01$
for the EAM,
$0.33 \pm 0.01$
for model I and
$0.31 \pm 0.01$
for model II barriers.
These results are lower than the anticipated
$\gamma \ge 0.4$ due to dimer (and trimer) mobility.
\end{abstract}


\twocolumn
\newpage
\narrowtext

\section{Introduction}

Recent experiments on thin film growth 
on well characterized substrates
using molecular beam epitaxy
have provided detailed information about growth kinetics and 
morphology. 
In particular, growth in the submonolayer regime has been studied
extensively using both scanning tunneling microscopy (STM) 
\cite{Mo91,Hwang91,Potschke91,Bott92,Michely93,Kopatzki93,Stroscio94,Gunther94,Esch94,Linderoth96}
and diffraction methods such as helium
atom beam scattering 
\cite{Ernst92,Li93,Ernst94}
and low energy electron diffraction
\cite{Zuo91,zuo94,Durr95}.
It was observed that for a variety of systems 
and a broad temperature range, island nucleation is
the dominant mechanism for crystal growth. 
A variety of island
morphologies has been found.
Fractal-like islands, 
resembling diffusion limited aggregation (DLA) clusters,
were observed 
in Au on Ru(0001) 
\cite{Hwang91,Potschke91},
Cu on Ru(0001)
\cite{Potschke91}
and Pt on Pt(111)
\cite{Bott92,Michely93}.
On the other hand, compact islands 
were observed
in Ni on Ni(001) system
\cite{Kopatzki93}. 
Scaling properties of the island density
as a function of deposition rate and coverage,
as well as the island size distribution
have been 
studied experimentally by several groups
\cite{Stroscio94,Gunther94,Linderoth96,Ernst92,Zuo91,zuo94}.
Field ion microscopy (FIM) experiments that 
provide direct access to
diffusion processes at the atomic scale, have also been performed
\cite{Kellogg90,Kellogg91,Kellogg94}.
This technique was used to identify the diffusion modes of 
adatoms
\cite{Kellogg90}
as well as
small islands 
\cite{Kellogg91,Kellogg94}
on FCC(001) metal surfaces
and to measure their diffusion
coefficients.

Theoretical studies aimed at providing a better understanding of
the relation between key processes at the atomic scale and the
resulting morphologies have been done using Monte Carlo (MC)
simulations
\cite{Voter86,Clarke88,Bartelt92,Bartelt93,Zhang93,Bales94,Ratsch94
,Barkema94,Zhang94,Schroeder95,Amar95,Ratsch95,Jacobsen95,Bales95,Amar96,Furman97}.
In simulations of island growth during deposition, 
atoms are deposited randomly on the substrate 
at rate $F$ [given in monolayer (ML) per second]
and then hop according 
to a microscopic model.
The hopping rate $h$ (in units of hops per second)
of a given atom to each unoccupied nearest neighbor (NN)
site is given by 
\begin{equation}
h=\nu \cdot \exp(-E_B/k_BT)
\label{hoppingrate}
\end{equation}
where 
$\nu = 10^{12}$ $s^{-1}$ 
is the commonly used 
attempt rate,
$E_B$ is the activation energy barrier, 
$k_B$ is the Boltzmann constant and $T$ is the temperature.

The activation energy barrier $E_B$ depends on the local environment of the
hopping atom, namely the configuration of occupied and
unoccupied adjacent sites.
Two approaches have been taken in the construction of
the energy barriers for hopping.
One approach was to construct simple models that include
the desired features, such as stability and mobility of 
small islands, and  that take into account properties such as bond
energies
\cite{Voter86,Clarke88,Bartelt92,Bartelt93,Bales94,Ratsch94
,Zhang94,Schroeder95,Amar95,Ratsch95,Bales95,Amar96,Furman97}.
The advantage of this approach is that the models are well
defined 
and use only a few parameters.
These models are useful for studies of scaling and
morphology but cannot provide a quantitative description
of diffusion on a particular substrate.
A second approach is based on the use of an approximate many-body 
energy functional to calculate the hopping energy
barriers for a complete set of relevant configurations
\cite{Zhang93,Barkema94,Jacobsen95}.
This approach provides a good description of diffusion
processes on the given substrate but only limited 
understanding due to the large number of parameters.

In this work we propose a procedure that combines the 
advantages of both approaches. 
Using sensible assumptions about the bond energies and 
diffusion paths we obtain a simple formula for the activation energy
barriers. 
We then optimize the parameters of this formula by using energy barriers 
obtained from the embedded atom method
for Cu diffusion on Cu(001).
This procedure provides two simple models
which combine the best features of both approaches.
Model I, which has three parameters, provides
good quantitative description of the landscape of
hopping energy barriers.
Model II, which has only two parameters and 
is as simple as other minimal models,
still 
incorporates the essential physics of 
diffusion of adatoms on FCC(001) metal surfaces.

We find that on the Cu(001) substrate, edge mobility 
(i.e., the mobility of an 
adatom at the perimeter of an island)
is much higher than the single adatom mobility,
giving rise to compact island shapes.
Dimer mobility is comparable to the single adatom mobility
while trimers are also mobile
\cite{Barkema94,Biham95,Boisvert97}.
Mobility of small islands has a significant effect on
the asymptotic scaling properties and is thus particularly
important.
We also find that vacancies have higher mobility than adatoms.

The paper is organized as follows. 
In Section II we introduce the models.
Simulation and results are presented
in Section III
with emphasis on scaling and morphology.
The results and their implications are discussed in
Section IV and a summary is given in Section V.

\section{The Models}

\subsection{The EAM Barriers}
In this work we
use a set of energy barriers for Cu on Cu(001) 
\cite{Karimi95}
obtained using  the embedded atom method (EAM)
\cite{Daw83}. 
This method uses 
semiempirical potentials and  provides a good 
description of
self diffusion of Cu on Cu(001) and similar surfaces
\cite{Karimi95}.
Specifically, we use the EAM
functions of Cu developed by Adams, Foiles, and Wolfer
\cite{Adams89}
which
are fitted to a similar data base as the one employed by Foiles,
Baskes, and Daw
\cite{Foiles86}.

The hopping energy barriers are calculated for all local environments
as shown in Fig. 1, where 
seven adjacent sites, $i=0,\dots,6$ are taken into account
\cite{3by3}.
Each one of these sites can be either 
occupied ($S_i=1$) or vacant ($S_i=0$), 
giving rise to $2^7=128$ barriers. 
A binary representation is used
to assign indices to these barriers. 
For each configuration $(S_0,\dots,S_6)$ the barrier is 
given by $E_B^n$,
where
\begin{equation}
n=\sum_{i=0}^6 S_i \cdot 2^i
\label{barindex}
\end{equation}
takes the values
$n=0,\dots,127$. 
The full set of hopping energy barriers (given in eV)
is presented in Table I.
To show these values in a compact form, 
each barrier in Table I corresponds to a
configuration in which the occupied sites are the union of the
occupied sites in the picture on top of the given column and
on the left hand side of the given row. 
The column in Table I in which a given configuration appears
is determined by the occupancy of sites
$i=2,3,6$
while the row is determined by sites
$i=0,1,4,5$.
One can define
\begin{equation}
n_1 = \sum_{i=2,3,6} S_i \cdot 2^i; \ \ \ n_2 = \sum_{i=0,1,4,5} S_i \cdot 2^i
\label{n1n2}
\end{equation}
such that for each configuration 
$n=n_1 + n_2$.
To demonstrate the use of Table I, we will check the barrier for the 
configuration in which sites $0,3$ and $4$ are occupied and all
other sites adjacent to the hopping atom are vacant.
For this configuration, according to Eq. 
(\ref{n1n2}),
$n_1=8$ and $n_2=17$ ($n=25$).
The barrier, which is found in the column with the
index $8$ and the row with index $17$,
is 
$E_B^{25} = 0.89$  eV.

In Table I we use the symmetries of the configurations in the
$3 \times 3$ cell to reduce the number of entries. 
There is a mirror symmetry plane perpendicular to the surface and 
containing the arrow of the 
hopping atom. 
Consequently, the columns of $n_1=4$ and $12$,
in which site $i=2$ is occupied, stand also for the 
symmetric configurations in which 
$i=6$ is occupied.
In the other four columns,
there are some configurations that, due to symmetry,
appear twice. 
In such cases, the barrier for the configuration with larger
$n$ appears in {\it italics}.

Here we consider only hopping moves in which a single atom  
hops each time.
However, it turns out that in some cases the molecular statics calculations,
used to obtain the barriers, give rise to concerted moves. 
In such moves the atom at site $i=3$ follows the hopping atom
and takes the  place vacated by the hopping atom. 
This fact significantly reduces the barrier.
We have found that for all configurations in which concerted
moves appear, these concerted moves can be suppressed by adding one more row
of atoms on the left hand sides of sites
$i=0,3$ and $4$.
In Table I, the energy values for those  configuration in which a concerted 
move was found, 
are shown in parenthesis. 
The barrier
obtained when the concerted move was suppressed 
is shown to the left of the parenthesis.

Concerted moves may affect film growth by modifying either the
stability and mobility of small islands or the morphology of
large islands.
The effect of the shear move
($n_1=8; \ n_2=3$),
in which concerted motion reduces the barrier from
$E_B^{11} = 0.78$ eV to $0.65$ eV, was studied in Ref.
\cite{Shi96}.
It was claimed that for experiments with
very low deposition rates this move effectively raises
$i^{\ast}$ 
from 3 to 8, 
where 
$i^{\ast}$ 
is the size of the critical island nucleus.
Simple calculations of the hopping rates,
using Eq.
(\ref{hoppingrate}) indicate that,
since the reduced barrier is still rather large,
the effect of this concerted move is negligible for the temperatures and 
deposition rates studied in this
paper .

We find that there are other concerted moves which, unlike
the shear move discussed above, cause a dramatic reduction
in the barrier for the given move (Table I).
If the atoms inside the $3 \times 3$ square represent an
isolated island and sites around are vacant,
it is easy to see that
moves such as 
$(n_1,n_2)=(34,12)$, and $(35,76)$
simply lead to a more stable configuration of the pentamer
and heptamer islands, respectively.
Once this more stable configuration is obtained, 
the concerted move cannot take place again.
Since these more stable configurations can be obtained through
other fast moves, the concerted move does not have a
significant effect on the overall morphology of the film.
Other moves such as 
$(34,8)$,
$(50,12)$,
and
$(51,76)$
can occur only as long as the islands of
4, 6 and 8 atoms respectively, have not
reached their most stable configuration.
If the above configurations, in which concerted moves are possible,
occur in a denser environment, and in particular 
if the sites on the left hand side of sites
$i=0,3$ and $4$ are occupied, the concerted
moves are suppressed.
Thus, we conclude that in both cases the concerted moves
do not have a significant effect in the 
simulations presented 
here and they will be ignored.

To gain a better understanding of the barrier energy 
landscape we discuss the
barrier height distribution 
(without concerted moves)
in Fig. 2(a). 
We observe that this distribution exhibits four peaks.
This feature is in agreement with Ref. 
\cite{Breeman94}
where a different method 
\cite{Finnis84}
was used to calculate the barriers. 
Each peak, corresponds to a single or a double column
in Table I (there is a little ambiguity in the region between
peaks III and IV due to  a slight overlap).
In general, peak I includes very fast moves toward island edges,
peak II includes moves along the edge, peak III includes, most notably,
the single atom move ,while peak IV includes detachment 
moves.

\subsection{Construction of Models}

When an atom on the surface hops into a vacant nearest neighbor site
it has to cross the energy barrier between the initial and final
sites. 
We have used molecular statics in conjunction with EAM functions
to find the diffusion path of an adatom. 
Not surprisingly, we have found that
the top of the energy barrier to go from a four-fold coordinate site to
an adjacent site is at the bridge site.
By slowly moving the adatom within the molecular statics calculation, it turns 
out that
the hopping energy barrier is simply the difference
between the energy at the bridge site and in the initial site. 
The occupancy of the seven adjacent sites 
(Fig. 1) affects both energies. 
We will now
express the energy of the hopping atom in its initial
site as:
\newline
\begin{eqnarray}
E_{in} & = & E_{in}^0 - \Delta E_{in} \cdot (S_1 + S_3 + S_5) \nonumber \\
       &   & - \Delta E_{in}^{nnn} \cdot (S_0 + S_2 + S_4 + S_6)
\end{eqnarray}
where $S_i=1$ if site $i$ is occupied and 0 otherwise. 
The energy of an isolated
atom is $E_{in}^0$ while
the reduction in its energy due the presence of an atom
in a nearest (next nearest) neighbor site is given by   
$\Delta E_{in}$ ($\Delta E_{in}^{nnn}$).
Here we assume that the contributions of nearest neighbor  
and next nearest neighbor (NNN) atoms to the energy are additive.
The energy of the hopping atom when it is 
on the bridge site is given by:  
\begin{equation}
E_{top} = E_{top}^0 - \Delta E_{top} \cdot (S_1 + S_2 + S_5 + S_6)
\end{equation}
where $E_{top}^0$ is the energy of an isolated 
atom on top of a bridge site, while
$\Delta E_{top}$ is the 
reduction
in the energy due to the presence of an
atom in one of the four sites adjacent to the bridge site. 
We do not include here NNN type contributions since their effect is small.
Therefore, for a given configuration the barrier 
$E_B = E_{top} - E_{in}$
for an atom to hop into an adjacent vacant site is given
by: 
\begin{eqnarray}
E_{B}^n & = & E_{B}^0 - \Delta E_{top}  \cdot (S_2 + S_6) 
		+ \Delta E_{in} \cdot S_3 \nonumber \\
        &   & - (\Delta E_{top} - \Delta E_{in}) \cdot (S_1 + S_5) \nonumber \\
	&   & + \Delta E_{in}^{nnn} \cdot (S_0 + S_2 + S_4 + S_6) 
\label{longform}
\end{eqnarray}
where
$E_B^0 = E_{top}^0 - E_{in}^0$
and $n$ is given by Eq. 
(\ref{barindex}).
To examine 
the formula above,
we  found the parameters that  best describe
the 128 EAM barriers by minimizing the sum of squares
\begin{equation}
R=\sum_{n=0}^{127} [E_B^n(EAM) - E_B^n(Eq. \ \ref{longform})]^2.
\label{optimize}
\end{equation}
The values obtained 
for the parameters in Eq. 
(\ref{longform})
are 
$E_B^0 = 0.494$, 
$\Delta E_{in} = 0.265$, 
$\Delta E_{top} = 0.268$
and 
$\Delta E_{in}^{nnn} = 0.024$ eV. 
Thus, we find that to within about $0.003$ eV,
$\Delta E_{in} = \Delta E_{top}$.
Replacing both   
$\Delta E_{in}$ 
and
$\Delta E_{top}$
by
$\Delta E = 0.265$ eV 
we obtain a three parameter model (model I) 
in which the energy barriers are:
\begin{eqnarray}
E_{B}^n & = & E_{B}^0 + \Delta E \cdot (S_3 - S_2 - S_6) \nonumber \\
        &   & + \Delta E_{in}^{nnn} \cdot (S_0 + S_2 + S_4 + S_6) 
\label{barrier3}
\end{eqnarray}
where $n$ is given by Eq.
(\ref{barindex}).
The distribution of energy barriers
obtained from Eq.  
(\ref{barrier3}) 
is shown in Fig. 2(b).
A very good agreement in the location of the four 
peaks with the EAM energy barriers is found, but
EAM peaks are significantly broader.
This can be due to the fact that during the hopping move
adjacent atoms can relax within their potential well. 
Model I accounts for these effects only on average and
therefore gives rise to narrower peaks.
 
One can further simplify the model by 
ignoring the NNN interactions which are relatively small, 
namely choosing 
$\Delta E_{in}^{nnn} = 0$.
This model (model II) has only two parameters 
and is described by:
\begin{equation}
E_{B}^n = E_{B}^0  
      +  \Delta E \cdot (S_3 - S_2 - S_6)
\label{barrier4}
\end{equation}
where $n$ is given by Eq.
(\ref{barindex}).
Repeating the optimization procedure of Eq.
(\ref{optimize})
for the barriers given by Eq.
(\ref{barrier4}),
we obtain
$E_B^0 = 0.526$ 
and 
$\Delta E = 0.255$.
In this model, all the energy barriers in each peak
collapse into a single value.
In spite of its simplicity, 
this model provides a good description 
of self diffusion of Cu on the Cu(001)
substrate.

\subsection{Diffusion Coefficients}

To find out which island sizes are mobile and to obtain 
the diffusion coefficients of mobile islands on
Cu(001),
we have done simulations of single cluster diffusion.
To obtain the statistics required for a precise 
determination of the diffusion
coefficients we performed 1000 runs
for diffusion of monomers, dimers, trimers, 
pentamers ($s=5$) 
and heptamers ($s=7$).
Each run was carried out for a  time equal to $1.0$ seconds, which is
more than 100
times larger than the time scale for hopping of an
isolated adatom
at the given temperature ($T=250$K). 
The diffusion coefficients  
were obtained from the relation
$\langle r^2 \rangle = 4 h_s t$, 
where $s=1,2,3,5,7$ is the number of atoms in the cluster,
$r$ is the distance between the initial
position of the center of mass of the
cluster and its position after 
time $t$, $h_s$ is the diffusion coefficient
for a cluster of size $s$ and 
$\langle ... \rangle$ represents an average over the 1000 runs.
The diffusion coefficients 
for monomers, dimers, trimers, pentamers and heptamers
are shown in
Table II.
It is also found
that under the conditions studied here
the diffusion coefficients for islands of size $s=4$, $6$
and $s \ge 8$ 
are negligible.

\section{Simulations and Results}

\subsection{Monte Carlo Simulations}

To assess the validity of our models, we have performed MC simulations of 
island growth
for a range of deposition rates
using the EAM  as well as model I and model II barriers
\cite{secondlayer}.
We used a continuous time MC technique 
\cite{Fichthorn91,Lu91,Clarke91,Kang94,Barkema94}
in which moves are selected randomly from the list of all possible moves
at the given time with the appropriate weights. 
The time is advanced after each move
according to the inverse of the sum of all rates.
The existence of four peaks in the
spectrum of energy barriers 
indicates that there are four typical time scales of hopping.
The two lowest peaks include very fast moves 
towards and along island edges
and motion of vacancies. 
The single atom move is in the
third peak while moves in which atoms detach from islands are
in the highest peak.

From statistics collected during the simulations we find that
most of the computer time is consumed by
moves $n=$6 (96) and 7 (112).
The first move occurs in trimers while both moves 
occur for atoms hopping along straight island edges. 
The reason that these moves consume so much time is
that 
the reverse move typically has the same low barrier
so the atom can continuously hop back and forth for a long time. 
To make the simulation feasible, we had to 
somewhat suppress these moves by
artificially raising
the barriers in the first and second peaks. 
This was done for all
barriers lower than 0.4 eV according
to:
$E_B^n \rightarrow E_B^n + \alpha (0.4 - E_B^n)$
where $\alpha = 0.7$. 
Since the moves associated with these barriers
are still orders of magnitude faster than for the two higher peaks
we expect that this modification will have only small effect on the
island morphology. 
We tried various values of $\alpha$ and found that
up to $\alpha \approx 0.8$ this is indeed the case.

\subsection{Island Growth and Morphology}

In Fig. 3 we show the island morphologies obtained in MC simulations
using the energy barriers from EAM [Fig. 3(a)], 
model I [Fig. 3(b)] and
model II [Fig. 3(c)].
These simulations were done on a
$250 \times 250$ lattice at 
$T=250$K and deposition rate
$F=0.01$ ML/s. 
The snapshots presented here are for a  coverage
$\theta=0.1$ ML.
These snapshots indicate that models I and II 
provide a good description of the model with the full EAM barriers 
as far as the island morphology is concerned.
The islands are rather compact, dominated by overall square
and rectangular shapes
with a small number of kinks.

The island density vs. coverage is presented in
Fig. 4 for the EAM ($\ast$), model I (o) and model II (+) energy barriers.
In all cases the island density quickly increases at low
coverage, then saturates and remains nearly constant thereafter.
It starts to decrease at higher coverage, 
when coalescence sets in.
For model I the island density is slightly higher than for the
EAM barriers while for model II it is about $60 \%$ higher. 
These differences are not intrinsic to the models.
They can
be traced to the fact that the 
single atom hopping energy barriers obtained from the
optimization procedure that we use 
are 
$E_B^0 = 0.494$ eV 
for model I and
$E_B^0 = 0.526$ eV
for model II; these values  are higher
than the value of
$E_B^0 = 0.485$ eV
from the EAM calculations.
These discrepancies could be fixed by an overall rescaling
of the barriers in each model so that the single atom 
barrier would be exactly equal to the EAM value. 
Since there is some arbitrariness in this procedure we 
chose not to apply it in the simulations presented here.
The important result of this work is that our models reproduce the 
essential features of diffusion, island growth
and scaling. 
The overall factor in island density 
to make the models
agree with the EAM calculations
could be obtained by rescaling.
While model I reproduces the spectrum of EAM
energy barriers more accurately,
the importance of model II is that it captures the essential features of 
adatom diffusion
on the Cu(001) in spite of its
simplicity .

Snapshots of the island morphology 
using the EAM barriers 
for three deposition rates,
$F=0.1$, 
$0.01$ 
and
$0.001$ ML/s
are presented in 
Fig. 5(a)-(c), respectively.
The simulations were done on a
$250 \times 250$ lattice at 
$T=250$K and
the snapshots were taken at
$\theta=0.1$ ML.
It is observed that decreasing the deposition rate gives rise to
fewer and larger islands.
In Fig. 6 we present the island density vs. coverage for the
EAM barriers, at deposition rates (from top to bottom)
$F=0.1, 0.04, 0.01, 0.004, 0.001$ and $0.0004$ ML/s.
It is observed that in addition to the decrease in island
density as the deposition rate decreases, 
the plateau becomes broader and flatter.

\subsection{Scaling Properties}

The island density vs. deposition rate is presented in 
Fig. 7 
for the EAM ($\ast$), model I (o) and model II (+)
energy barriers for
$T=250$K and
$\theta=0.1$ ML.
This coverage is well within the quasi-steady state
regime in which the island density is nearly constant
(see Fig. 6).
In this regime scaling arguments and rate equations 
indicate that in the asymptotic limit of low deposition rate
the island density exhibits power law behavior of the form

\begin{equation}
N \sim \left( {F \over {4 h_1}} \right)^{\gamma}
\end{equation}
where $h_1$ is the adatom diffusion coefficient
(the hopping rate for an isolated atom is $4 h_1$ due
to the four possible directions for hopping).
The exponent $\gamma$ depends on microscopic properties
of the system such as stability 
\cite{Stoyanov81,Venables84,Tang93,Amar95,Amar96,Bartelt95}
and mobility 
\cite{Villain92,Furman97}
of small islands,
anisotropic diffusion
\cite{Mo91,Linderoth96}
and magic islands
(islands which are stable while larger islands are unstable)
\cite{Schroeder95}.
It was found that for systems in which the smallest stable island
is of size $i^\ast + 1$ 
(where islands of size $s \le i^\ast$ dissociate),
$\gamma = i^\ast / (i^\ast + 2)$.
In this case
$i^{\ast}$ is the critical island size
\cite{Stoyanov81}.

Under the conditions studied here, detachment of atoms from 
islands is negligible, while mobility of small islands is
a significant effect.
Therefore, we define here the critical island size $i^{\ast}$
as the size for
which all islands of size 
$s \le i^{\ast}$ 
are mobile while
islands of size 
$s \ge i^{\ast}+1$ 
are immobile.
In this case
the asymptotic value of 
$\gamma$ 
in the limit  
where $F/h_1 \rightarrow 0$ 
is given by
\cite{Villain92,Furman97}:
\begin{equation}
\gamma = \frac{i^\ast}{2 i^\ast + 1}.
\label{eq:gamma}
\end{equation}
This result is exact if all mobile islands have the same diffusion
coefficient, 
namely $h_s = h$, $s = 1,\ldots,i^\ast$.
Still, it is a good approximation
if all the diffusion coefficients are of the same order of magnitude.
Table II shows that the diffusion coefficients of pentamers and
heptamers are practically negligible.
For EAM the diffusion coefficient of the trimer is also rather
small and we expect 
the system
to be described by $i^{\ast}=2$.
Models I and II exhibit small but not negligible mobility of trimers
and therefore we expect them to be described by either
$i^{\ast}=2$ or
$i^{\ast}=3$.

In Fig. 7 we observe nearly two and a half decades of scaling behavior, 
in a range of deposition rates between
$F=0.1$ and $0.0004$ ML/s.
The best fit through the six data points 
provides
$\gamma=0.32 \pm 0.01$
for the EAM,
$0.33 \pm 0.01$
for model I and
$0.31 \pm 0.01$
for model II barriers.
These results are significantly lower than the asymptotic values for
$i^{\ast}=2$ ($\gamma = 2/5$) or
$i^{\ast}=3$ ($\gamma = 3/7$).
In Ref.\
	\cite{Furman97}
a similar deviation was found between simulation results
and the asymptotic predictions of the rate equations.
It was argued that this deviation is due to the slow convergence 
of $\gamma$ to its  
asymptotic value, 
when the deposition rate $F$ is lowered.
This conclusion is strengthened by the fact that
if only the four left-most data points are used in the linear
fit,
the resulting $\gamma$ equals $0.35 \pm 0.01$
for the EAM,
$0.36 \pm 0.01$
for model I and
$0.34 \pm 0.01$
for model II.
These results are
somewhat closer to the asymptotic value.
Moreover, within the error bars they are larger than
the asymptotic value that arises from adatom mobility alone,
namely $\gamma=1/3$.
Therefore, the possibility that
$\gamma$ converges to $1/3$ 
should be discarded.
Since the conditions used here are experimentally relevant, our
results indicate that one should be careful in drawing conclusions
about processes at the atomic scale from scaling results
(see Section IV D).

The scaling properties of the island size distribution have
been studied both experimentally 
\cite{Zuo91,Stroscio94}
and theoretically
\cite{Bartelt92,Bartelt93,Schroeder95,Amar95,Amar96,Furman97}.
These studies indicated that the island size distribution
depends on the stability and mobility of small islands
and is modified in the case of magic island sizes
\cite{Schroeder95}.
The scaled island size distributions 
are presented
for the three models in Fig. 8.
For all three models the deposition rates are
$F=10^{-1}$ ($\ast$),
$10^{-2}$ (o)
and
$10^{-3}$ ML/s (+).
These results are based on statistics collected from
20 runs on a 
$250 \times 250$ size lattice
at $T=250$K and
$\theta=0.1$ ML.

The general shape of the distributions seems to resemble 
previous results 
\cite{Furman97}
for 
$i^{\ast}=2,3$ 
where the peaks rise more slowly on the left hand size
(small $s/\bar s$) due to the depletion of the mobile islands.
This trend is qualitatively similar to the results for
the case where small islands are unstable, as
shown in Ref.
\cite{Amar95,Amar96}.
Note also that the peak height increases considerably as $F$ 
decreases.
This may be due to coalescence which is found to become more
pronounced as the deposition rate decreases.
Coalescence causes $\bar s$ to increase, pushing up the
scaled island size distribution 
which includes the factor
${\bar s}^2/\theta$.

\section{Discussion}

\subsection{Small-Island Mobility} 
For both the EAM barriers and models
I and II 
we obtain significant dimer mobility.
The trimer mobility for the EAM barriers is rather small,
and somewhat larger in models I and II.
In general, we find that the energy barriers 
relevant for dimer and trimer mobility are in the same
peak as the single atom hopping. 
The differences in hopping rates between monomers, dimers 
and trimers are due to very small differences in the
energy barriers, which at $T=250$K 
are significant.
These differences 
in hopping rates
decrease as the temperature increases.
In model II, which has only one activation barrier for 
each peak, the activation barriers for monomer, dimer and
trimer mobilities are all equal.
Combinatorial summation of paths then shows that for model II
the diffusion coefficient of dimers (trimers)
is equal to
one half (one quarter) of the monomer diffusion
coefficient, at all temperatures.
Therefore, in model II, at any temperature in which 
atoms are mobile, dimers and trimers are also mobile. 
The mobility of small islands has a significant effect on 
the asymptotic scaling of the island density
vs. deposition rate.
In principle, this can be used to extract these diffusion
properties at the atomic scale from STM and diffraction
results at larger scales.
However, our simulations indicate that for experimentally
relevant parameters the system is likely to be away from
the asymptotic regime, 
and therefore one should be careful in drawing conclusions
from the empirically determined scaling exponents.
The relation between mobility of small islands and edge mobility
was examined
in Ref.
\cite{Furman97}.
It was shown that both types of mobilities
are generated by the same hopping moves and are therefore related,
namely, small island mobility implies edge mobility.

\subsection{Edge Mobility}

We have found that for Cu(001) atom mobility along
straight island edges (for which the barriers belong to
peak II) is much  
higher than the single atom mobility
(with a barrier in peak III).
This indicates, using an argument from Ref. 
\cite{Zhang94}, 
that islands should form compact shapes
for a broad range of temperatures. 
This conclusion is confirmed by our MC simulations and is 
in agreement with STM results
\cite{Kopatzki93}. 
For the temperature studied here detachment of atoms from
islands is negligible
and therefore edge mobility is the dominant process which
shapes the islands.

\subsection{Vacancy Mobility} 

It is found that for Cu(001) the mobility of a single vacancy
is higher than the single adatom mobility. 
We observe that the barrier for the diffusion of a single
vacancy is sensitive to the environment beyond the 
$3 \times 3$ 
square used to calculate energies in this paper.
The 
energy barrier 
for diffusion of a single vacancy is
$E_B^{127}=0.12$ eV 
if all sites around the 
$3 \times 3$ 
square are vacant. 
Adding three more atoms on the left it increases it to 
$0.28$ eV.
Embedding the hopping vacancy in an occupied 
$5 \times 5$
square increases the barrier to
$0.31$ eV 
while the barrier for a single vacancy in the substrate
is found to be
$0.43$ eV
(using a slab of 20 layers, 121 atoms/layer).
We conclude that, although the vacancy mobility is higher than
the adatom mobility, at high coverage the difference may not be
dramatic
\cite{Boisvert97}. 
For example, at low temperature  a difference
of $0.05$ eV in the barrier may significantly affect the hopping
rate.
A more general conclusion is that models I and II best
describe diffusion and nucleation at relatively low coverage.
At high coverage, far beyond the percolation threshold 
hopping can be considered primarily as motion of vacancies.
When the adlayer is already dense, relaxation effects during the
moves are more important than at low coverage, and the 
configuration of occupied sites beyond the 
$3 \times 3$ square used here may affect the energy barriers. 
In spite of the prediction that a single vacancy is more mobile
than an adatom, 
note that for Ag(001) there is evidence that vacancy clusters
have lower diffusivity than islands
\cite{Wen96}.

\subsection{Comparison with Experiments}

The simulations presented here were carried out
with parameters typical to homoepitaxial growth experiments
on Cu(001) surfaces, namely,
linear lattice size of 250 sites (about 650 \AA),
$T = 250$K and
$F$ between $4 \times 10^{-4}$ and $10^{-1}$ ML/s
\cite{Ernst92,zuo94}.
In Ref.\ 
\cite{zuo94}.
$\gamma = 0.33$ was measured for $T = 223$K,
$F$ between $2.5 \times 10^{-4}$ and $10^{-3}$ ML/s,
and coverage $\theta = 0.3$.
This value of $\gamma$ 
is in a very good agreement with our simulation results.
The authors of Ref.\ 
\cite{zuo94}
claimed that this indicates that
in this system and parameter range, $i^\ast = 1$.
Our results indicate that in this parameter range
the system is away from the asymptotic regime.
Thus, the value $\gamma = 0.33$ can be obtained
eventhough $i^\ast = 2$ or $3$.
A value of $\gamma = 0.55$ was reported 
(for $T = 223$K, and $F$ between $10^{-3}$ and $5 \times 10^{-3}$)
and $\gamma = 0.58$
(for $T = 263$K, and $F$ between $2.5 \times 10^{-4}$ and $10^{-3}$)
in the same coverage, $\theta = 0.3$
\cite{zuo94}.
These values are close to those reported in Ref.\ 
\cite{Ernst92}, namely
$\gamma = 0.54$
(for $T = 220$K,
and $F$ between $6.7 \times 10^{-4}$ and $3.3 \times 10^{-3}$),
and $\gamma = 0.46$
(for $T = 230$K,
and $F$ between $4 \times 10^{-4}$ and $10^{-2}$).
A possible interpretation of the higher values of $\gamma$
observed in these experiments,
is that processes such as
coalescence become significant as the coverage increases.

\subsection{Diffusion on Other FCC(001) Metal Surfaces}

We expect our approach to apply also to other FCC(001) metal
surfaces such as Ni(001) and Ag(001) in which diffusion occurs through
hopping rather than  by the exchange mechanism. 
EAM calculations for Ni(001) 
and Ag(001) are consistent with our
models
\cite{NiAg}.
For surfaces in which the exchange mechanism is favorable, as it is
believed to be the case 
for Al
\cite{Feibelman90}, 
Pt
\cite{Kellogg90}, 
Pd and Au 
\cite{Liu91}
modifications
of the models are required.
However, we believe that the approach presented here should still
provide a simple and useful model for diffusion on these surfaces.

\section{Summary}

We have studied adatom self diffusion and island growth on Cu(001)
using MC simulations at the atomic scale.
As input to  the MC simulation  we used a complete set of
energy barriers obtained from the embedded atom method.
To reduce the number of parameters and to obtain better
understanding of the diffusion processes we have
constructed two simple models.
Model I, with three parameters,
provides a good quantitative description of the full landscape 
of hopping energy barriers.
Model II, with only two parameters, is a minimal model
which still incorporates the essential features of adatom 
diffusion on the Cu(001) surface.

We examined the diffusion properties of small islands and 
found that the mobility of dimers is comparable to the
single adatom mobility while trimers are also somewhat
mobile. 
The mobility of adatoms along island edges was found to be
much higher than the single adatom mobility.
Since atoms detachment from islands is negligible, 
edge mobility is the dominant process that shapes 
islands into the compact forms.
A further conclusion from the EAM calculations is that vacancy
mobility, which is dominant at very high coverages, is higher
than the single adatom mobility, which is dominant at low
and intermediate coverages.

MC simulations of island growth show similar morphologies
for the EAM barriers and models I and II.
In all cases the islands generally form square or rectangular
shapes with a small number of kinks.
Studies of the scaling of the island density $N$ 
vs. deposition rate $F$ show that
$N \sim F^{\gamma}$ 
where 
$\gamma=0.32 \pm 0.01$
for the EAM,
$0.33 \pm 0.01$
for model I and
$0.31 \pm 0.01$
for model II barriers.
These results are lower than the asymptotic value 
of 
$\gamma \ge 0.4$
anticipated for
systems that exhibit dimer (and even trimer)
mobility.
It indicates that for the experimentally relevant 
conditions studied here the system has not reached
the asymptotic regime.
This result call for caution in using scaling results
to draw conclusions about diffusion properties at the
atomic scale.

Since both models I and II have only few parameters, one needs a very small
set of calculated energy barriers to determine these parameters 
for a given FCC(001) substrate, and a few more barriers to verify
that the model applies to that substrate.
This may open the way for a more effective use of first principle 
calculations of energy barriers on the surface as an input
to the kinetics calculations. 
We believe that the procedure described in this paper would prove
useful much beyond the FCC(001) metal substrates.
We expect it to
be applicable with some modifications to take into account exchange moves
which are believed to be 
favorable in
Al,Pt,Pd and Au.
We believe that the approach presented here will provide useful models 
for adatom self diffusion on
FCC(111) 
surfaces, for which reliable calculations of energy barriers 
are becoming available
\cite{Stumpf94,Stumpf96},
and also to a variety of 
heteroepitaxial systems.

\acknowledgments

OB would like to thank Oded Millo and Nira Shimoni 
for helpful discussions.
We would like to acknowledge partial support from the 
NSF under grants DMR-9217284 and DMR-9119735 at Syracuse
University during the initial stages of this work and
partial support from Intel-Israel college relations committee
during the final stages.

\newpage
\begin{center}
	{\large Table Captions}
\end{center}

\begin{table}
\caption{
The hopping energy barriers 
obtained from the EAM calculations
for all 128 possible configurations
within a $3 \times 3$ square around the hopping atom. 
The barriers
are given in eV.
Each number in the Table is the barrier $E_B^n$ for the
configuration in which the occupied sites are the union of the
occupied sites in the picture on top of the given column 
(indexed by $n_1$)
and on the left hand side of the given row
(indexed by $n_2$). 
Consequently, the index 
$n$ specifying the barrier is given by $n=n_1 + n_2$.
}
\label{tab:barriers}
\end{table}

\begin{table}
\caption{
Diffusion coefficients of small islands 
of sizes $s=1,2,3,5,7$ atoms
as measured
from MC simulations of single islands.
Results are shown for the 
EAM, model I and model II barriers
at $T=250$K.
Islands of sizes $s=4$, $6$
and $s \ge 8$ were found to be 
immobile at this temperature.
}
\label{tab:diff.coef}
\end{table}

\onecolumn

\setcounter{table}{0}
\newcommand{\move}{
        \put(0,110){\line(1,0){30}}
        \put(0,120){\line(1,0){30}}
        \put(0,130){\line(1,0){30}}
        \put(0,140){\line(1,0){30}}
        \put(0,140){\line(0,-1){30}}
        \put(10,140){\line(0,-1){30}}
        \put(20,140){\line(0,-1){30}}
        \put(30,140){\line(0,-1){30}}
        \put(15,125){\circle{7}}
        \put(19,125){\vector(1,0){6}}
}
\newcommand{\lzero}{
\begin{picture}(25,25)(-15,92)
        \move
\end{picture}
}
\newcommand{\lone}{
\begin{picture}(25,25)(-15,92)
        \move
        \put(5,135){\circle{7}}
\end{picture}
}
\newcommand{\ltwo}{
\begin{picture}(25,25)(-15,92)
        \move
        \put(15,135){\circle{7}}
\end{picture}
}
\newcommand{\lthree}{
\begin{picture}(25,25)(-15,92)
        \move
        \put(5,135){\circle{7}}
        \put(15,135){\circle{7}}
\end{picture}
}
\newcommand{\lsixteen}{
\begin{picture}(25,25)(-15,92)
        \move
        \put(5,115){\circle{7}}
\end{picture}
}
\newcommand{\lseventeen}{
\begin{picture}(25,25)(-15,92)
        \move
        \put(5,135){\circle{7}}
        \put(5,115){\circle{7}}
\end{picture}
}
\newcommand{\leighteen}{
\begin{picture}(25,25)(-15,92)
        \move
        \put(15,135){\circle{7}}
        \put(5,115){\circle{7}}
\end{picture}
}
\newcommand{\lnineteen}{
\begin{picture}(25,25)(-15,92)
        \move
        \put(5,135){\circle{7}}
        \put(15,135){\circle{7}}
        \put(5,115){\circle{7}}
\end{picture}
}
\newcommand{\lthirtytwo}{
\begin{picture}(25,25)(-15,92)
        \move
        \put(15,115){\circle{7}}
\end{picture}
}
\newcommand{\lthirtythree}{
\begin{picture}(25,25)(-15,92)
        \move
        \put(5,135){\circle{7}}
        \put(15,115){\circle{7}}
\end{picture}
}
\newcommand{\lthirtyfour}{
\begin{picture}(25,25)(-15,92)
        \move
        \put(15,135){\circle{7}}
        \put(15,115){\circle{7}}
\end{picture}
}
\newcommand{\lthirtyfive}{
\begin{picture}(25,25)(-15,92)
        \move
        \put(5,135){\circle{7}}
        \put(15,135){\circle{7}}
        \put(15,115){\circle{7}}
\end{picture}
}
\newcommand{\lfortyeight}{
\begin{picture}(25,25)(-15,92)
        \move
        \put(5,115){\circle{7}}
        \put(15,115){\circle{7}}
\end{picture}
}
\newcommand{\lfortynine}{
\begin{picture}(25,25)(-15,92)
        \move
        \put(5,135){\circle{7}}
        \put(5,115){\circle{7}}
        \put(15,115){\circle{7}}
\end{picture}
}
\newcommand{\lfifty}{
\begin{picture}(25,25)(-15,92)
        \move
        \put(15,135){\circle{7}}
        \put(5,115){\circle{7}}
        \put(15,115){\circle{7}}
\end{picture}
}
\newcommand{\lfiftyone}{
\begin{picture}(25,25)(-15,92)
        \move
        \put(5,135){\circle{7}}
        \put(15,135){\circle{7}}
        \put(5,115){\circle{7}}
        \put(15,115){\circle{7}}
\end{picture}
}
\newcommand{\tsixtyeight}{
\begin{picture}(25,25)(2,110)
        \move
        \put(25,135){\circle{7}}
        \put(25,115){\circle{7}}
\end{picture}
}
\newcommand{\tseventysix}{
\begin{picture}(25,25)(2,110)
        \move
        \put(25,135){\circle{7}}
        \put(25,115){\circle{7}}
        \put(5,125){\circle{7}}
\end{picture}
}
\newcommand{\tfour}{
\begin{picture}(25,25)(2,110)
        \move
        \put(25,135){\circle{7}}
\end{picture}
}
\newcommand{\ttwelve}{
\begin{picture}(25,25)(2,110)
        \move
        \put(25,135){\circle{7}}
        \put(5,125){\circle{7}}
\end{picture}
}
\newcommand{\tzero}{
\begin{picture}(25,25)(2,110)
        \move
\end{picture}
}
\newcommand{\teight}{
\begin{picture}(25,25)(2,110)
        \move
        \put(5,125){\circle{7}}
\end{picture}
}
\newcommand{\narrow}{
\begin{picture}(25,25)(2,110)
        \put(5,125){$\downarrow$}
\end{picture}
}
\begin{center}
\large{Table I}
\end{center}
\begin{center}
\begin{table}
\begin{tabular}{|l||c|cc|cc|c|} \hline 
                 &  Peak I& Peak   &   II   & Peak   &   III  & Peak IV \\ \hline 
\hspace{0.2in} $n_1 \rightarrow$ \hspace{0.2in}   &   68   &   76   &   4    &   12   &   0    &   8    \\ 
   $n_2$         & \hspace{0.5in}  & \hspace{0.5in}   & \hspace{0.5in}  & \hspace{0.5in}  &  \hspace{0.5in}  & \hspace{0.5in}\\ 
 \narrow         & \tsixtyeight       & \tseventysix       & \tfour       & \ttwelve       & \tzero       & \teight    \\ \hline \hline
            &     &        &        &        &        &        \\ 
            &     &        &        &        &        &        \\ 
  0    &  0.01  &  0.25  &  0.18  &  0.48  &  0.48  &  0.81  \\ 
  \lzero          &        &        &        &        &        &        \\ 
  1         &  0.02  &  0.28  &  0.25  &  0.53  &  0.46  &  0.85  \\ 
  \lone          &        &        &        &        &        &        \\ 
  2    &  0.02  &  0.21  &  0.18  &  0.44 (0.34)  &  0.46  &  0.74 (0.60) \\ 
  \ltwo    &        &        &        &        &        &        \\ 
  3   &  0.05  &  0.25  &  0.24  &  0.48  &  0.54  &  0.78 (0.65)  \\ 
  \lthree    &        &        &        &        &        &        \\ 
  16  &  {\it 0.02 }     & {\it 0.28 }   &  0.21  &  0.50  & {\it 0.46}   & {\it 0.85}    \\ 
  \lsixteen    &        &        &        &        &        &        \\ 
  17  &  0.04  &  0.30  &  0.28  &  0.54  &  0.66  &  0.89  \\ 
  \lseventeen    &        &        &        &        &        &        \\ 
  18  &  0.05  &  0.27  &  0.23  &  0.49  &  0.52  &  0.80  \\ 
  \leighteen    &        &        &        &        &        &        \\ 
  19  &  0.08  &  0.29  &  0.29  &  0.52  &  0.61  &  0.83  \\ 
  \lnineteen    &        &        &        &        &        &        \\ 
  32  &   {\it 0.02}     &   {\it 0.21}     &  0.18  &  0.48  &   {\it 0.46}     & {\it 0.74 (0.60) }     \\ 
  \lthirtytwo    &        &        &        &        &        &        \\ 
  33  & {\it 0.05 }  & {\it 0.27 }   &  0.28  &  0.54  & {\it 0.52 } & {\it 0.80 }  \\ 
  \lthirtythree  &        &        &        &        &        &        \\ 
  34  &  0.05  &  0.24  &  0.28  &  0.50 (0.16)  &  0.55  &  0.78 (0.37)  \\ 
  \lthirtyfour    &        &        &        &        &        &        \\ 
  35  &  0.08  &  0.26 (0.08)  &  0.33  &  0.49  &  0.62  &  0.81 (0.47)  \\ 
  \lthirtyfive    &        &        &        &        &        &        \\ 
  48  &  {\it 0.05} & {\it 0.25}    &  0.28  &  0.51  & {\it 0.54}   &   {\it 0.78 (0.65)}     \\ 
  \lfortyeight    &        &        &        &        &        &        \\ 
  49  &  {\it 0.08}      &  {\it 0.29}     &  0.33  &  0.56  &   {\it 0.61}     &   {\it 0.83}  \\ 
  \lfortynine    &        &        &        &        &        &        \\ 
  50  &   {\it 0.08}     &   {\it 0.26 (0.08)}     &  0.34  &  0.51 (0.22)  &   {\it 0.62}     &   {\it 0.81 (0.47)}     \\ 
  \lfifty    &        &        &        &        &        &        \\ 
  51  &  0.13  &  0.28 (0.12) &  0.40  &  0.53 (0.36)  &  0.70  &  0.69  \\ 
  \lfiftyone   &        &        &        &        &        &        \\ \hline
\end{tabular}
\end{table}
\end{center}

\begin{center}
\large{Table II}
\end{center}

\begin{table}
\vskip-\lastskip
\begin{tabular}{cr@{}l@{${}\pm{}$}r@{}l
			r@{}l@{${}\pm{}$}r@{}l
				r@{}l@{${}\pm{}$}r@{}l}
	Cluster & \multicolumn{12}{c}{Diffusion Coefficients
					[hops/s]}
	\\ 	\cline{2-13}
	Size & \multicolumn{4}{c}{EAM} &
		\multicolumn{4}{c}{model 1} &
		\multicolumn{4}{c}{model 2} 
	\\	\hline

	1 &  & 172 & 5 &  &  & 118 & 4  &  &  & 24.3  & 0  & .8
	\\
	2 &  & 237 & 8 &  &  & 55  & 2  &  &  & 13.2  & 0  & .4
	\\
	3 &  & 5.6 & 0 & .2 &  & 12.4 & 0 & .4 &  & 5.9 & 0  & .2
	\\
	5 &  & 1.12 & 0 & .04 &  & 0.94 & 0 & .03 &  & 1.00 & 0 & .03
	\\
	7 &  & 0.28 & 0 & .02 &  & 0.34 & 0 & .01 &  & 0.42 & 0 & .02
	\\
	\end{tabular}
\end{table}

\begin{center}
	{\large Figure Captions}
\end{center}

\figure{Fig. 1: 
Classification of all possible local environments 
of a hopping atom including seven adjacent sites. 
Each site can
be either occupied or unoccupied, 
giving rise to $2^{7}=128$ local 
environments. 
Sites 1, 3 and 5 are nearest neighbors of the original 
site while sites 1, 2, 5 and 6 are adjacent to the bridge site 
that the atom has to pass.
\label{fig1} }

\figure{Fig. 2:
The distribution of hopping energy barriers obtained from EAM calculations
(a) and from model I (b) for the 128 possible local environments of Fig. 1.
According to model I the lowest and highest peaks 
include 16 barriers each, while the middle peaks include 
48 barriers each. The EAM peaks are broad and some overlap occurs.
\label{fig2} }

\figure{Fig. 3: 
Snapshots of surface morphologies obtained in MC  
simulations using the EAM  
(a),
model I (b),
and 
model II (c) barriers.
The simulations were carried out on a
$250 \times 250$ lattice at 
$T=250$K and deposition rate
$F=0.01$ ML/s. 
The snapshots presented are at coverage
$\theta=0.1$ ML.
In all three cases
the islands form generally compact square and rectangular
shapes with a small number of kinks.
\label{fig3} }

\figure{Fig. 4:
The island density vs. coverage is presented for the EAM ($\ast$), model I (o) 
and model II (+) barriers.
In all cases the island density quickly increases at low
coverage. 
It then saturates and remains nearly constant
within the quasi-steady state aggregation dominated regime.
The solid line is a guide to the eye.
\label{fig4} }

\figure{Fig. 5: 
Snapshots of the island morphology 
obtained from MC simulations using 
the EAM barriers 
for three deposition rates:
(a) $0.1$; 
(b) $0.01$ 
and
(c) $0.001$ ML/s. 
The simulations were done on a
$250 \times 250$ lattice at 
$T=250$K and
the snapshots were taken at
$\theta=0.1$ ML.
It is observed that decreasing the deposition rate gives rise to
fewer and larger islands.
\label{fig5} }

\figure{Fig. 6: 
The island density vs. coverage is presented for the
EAM barriers, at deposition rates (from top to bottom)
$F=0.1, 0.04, 0.01, 0.004, 0.001$ and $0.0004$ ML/s.
The island density decreases as the deposition rate
is lowered.
Also, the plateau associated with the aggregation
dominated regime broadens.
The solid line is a guide to the eye.
\label{fig6} }

\figure{Fig. 7: 
The island density vs. deposition rate 
for the EAM ($\ast$), model I (o) and model II (+) barriers.
These results were obtained at
$T=250$K and $\theta=0.1$ ML
for deposition rates between
$F=0.1$ and $0.0004$ ML/s.
The best fit through the six data points in each of the
three models gives
$\gamma=0.32 \pm 0.01$
for the EAM,
$0.33 \pm 0.01$
for model I and
$0.31 \pm 0.01$
for model II barriers.
These results
indicate that models I and II provide a good description
of the scaling properties.
\label{fig7} }

\figure{Fig. 8: 
Scaled island size distributions are shown
for the EAM 
(a), 
model I 
(b), and 
model II
(c) barriers. 
$\bar s$ is the average island size.
The deposition rates examined in each case are
$F=10^{-1}$ ($\ast$),
$10^{-2}$ (o)
and
$10^{-3}$ ML/s (+).
These results are based on
20 runs on a 
$250 \times 250$ size lattice
at $T=250$K and
$\theta=0.1$ ML.
\label{fig8} }


\begin{references}
\bibitem{3by3}    
We reduced the neighborhood to a $3 \times 3$ square since we found
that the three extra sites used in Ref. 
\cite{Barkema94}
have little effect.
\bibitem{secondlayer} 
In experiments, no significant nucleation was observed
on top of islands up to nearly full coverage
\cite{Kopatzki93,Ernst92}. 
Therefore, in our simulations an atom 
deposited on top of an island hops randomly
until it jumps down to the lower layer.
\bibitem{NiAg}
To demonstrate that our models are applicable to
Ni(001) and Ag(001) we present here the results
of EAM calculations for the six configurations in
the first row of Table I.
For Ni(001) these barriers are
$E_B^n = 0.06$, $0.35$, $0.25$, $0.62$, $0.63$ and  $1.02$ eV
for
$n= 68, 76, 4, 12, 0$ and $8$ 
respectively.
For Ag(001) these barriers are
$E_B^n = 0.14$, $0.35$, $0.21$, $0.46$, $0.47$ and $0.71$ eV
for
$n= 68, 76, 4, 12, 0$ and $8$ 
respectively.
We observe that in both cases barriers which belong
to different peaks are typically separated by
$0.2$eV or more.
\end{references}
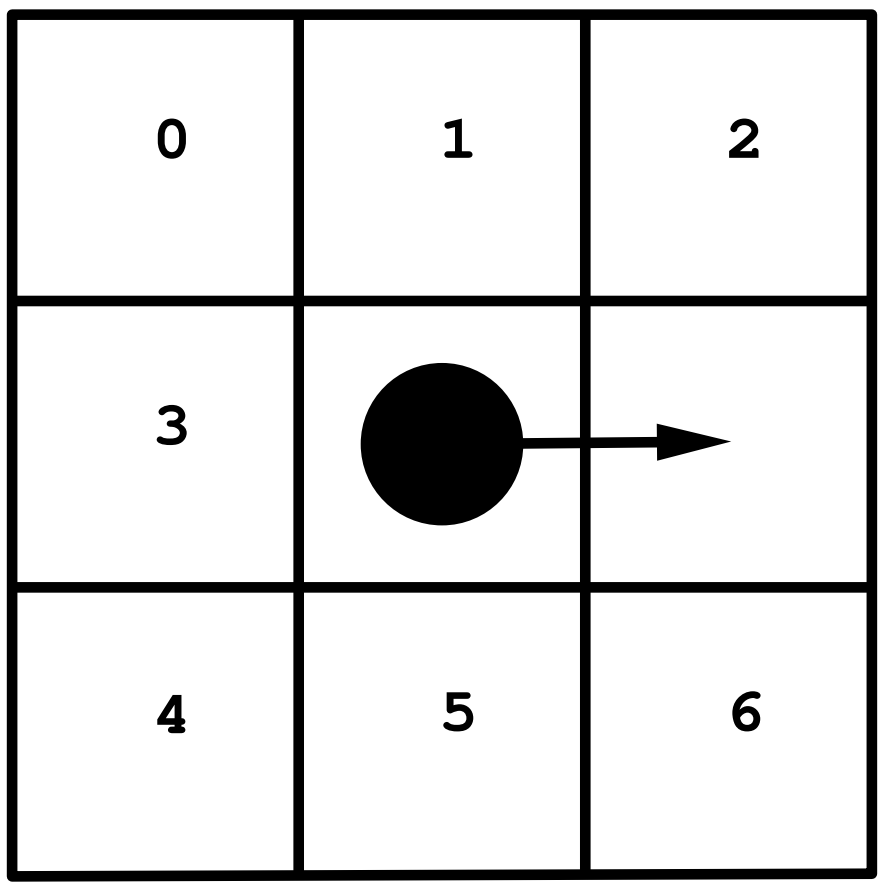
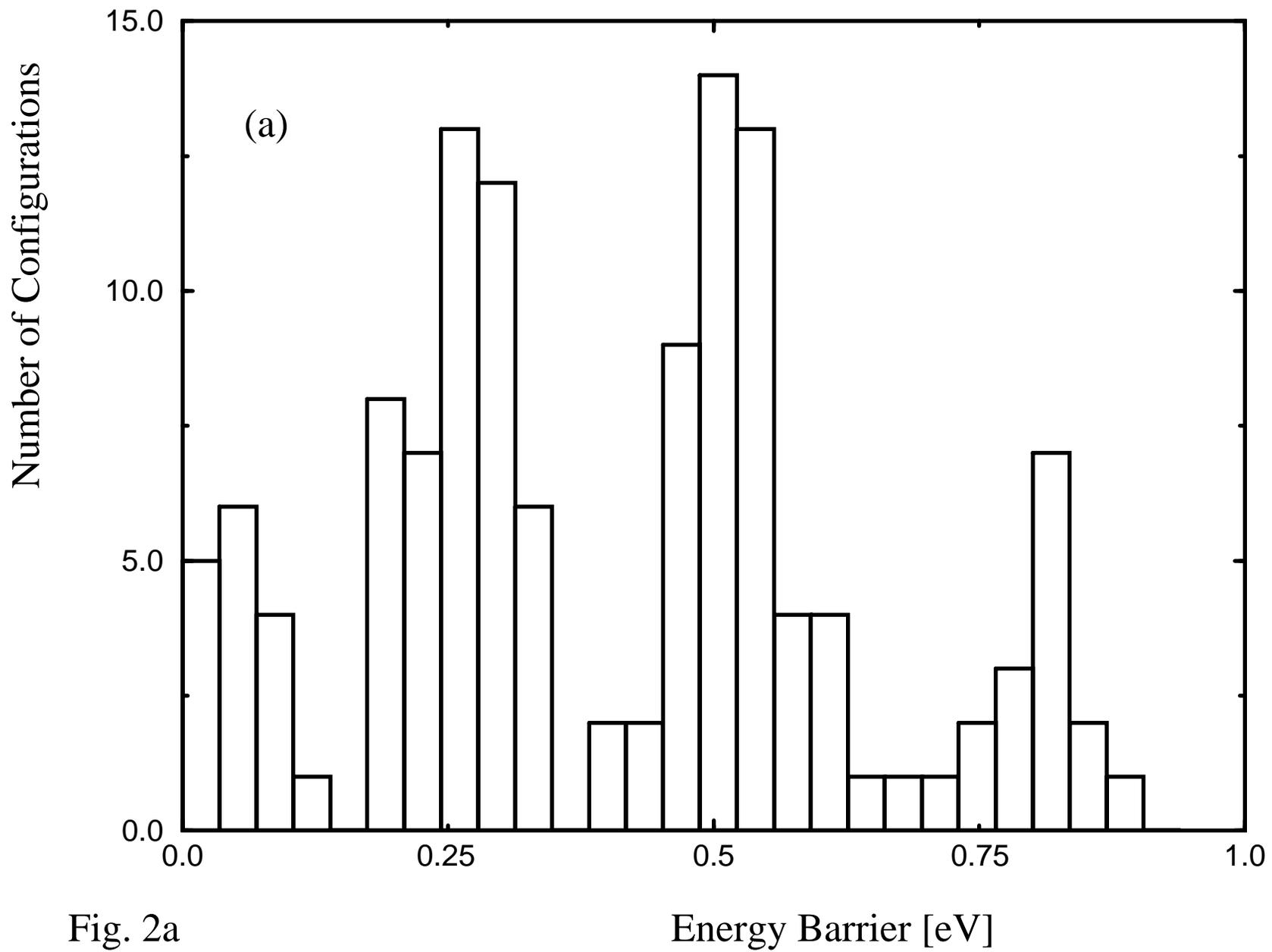
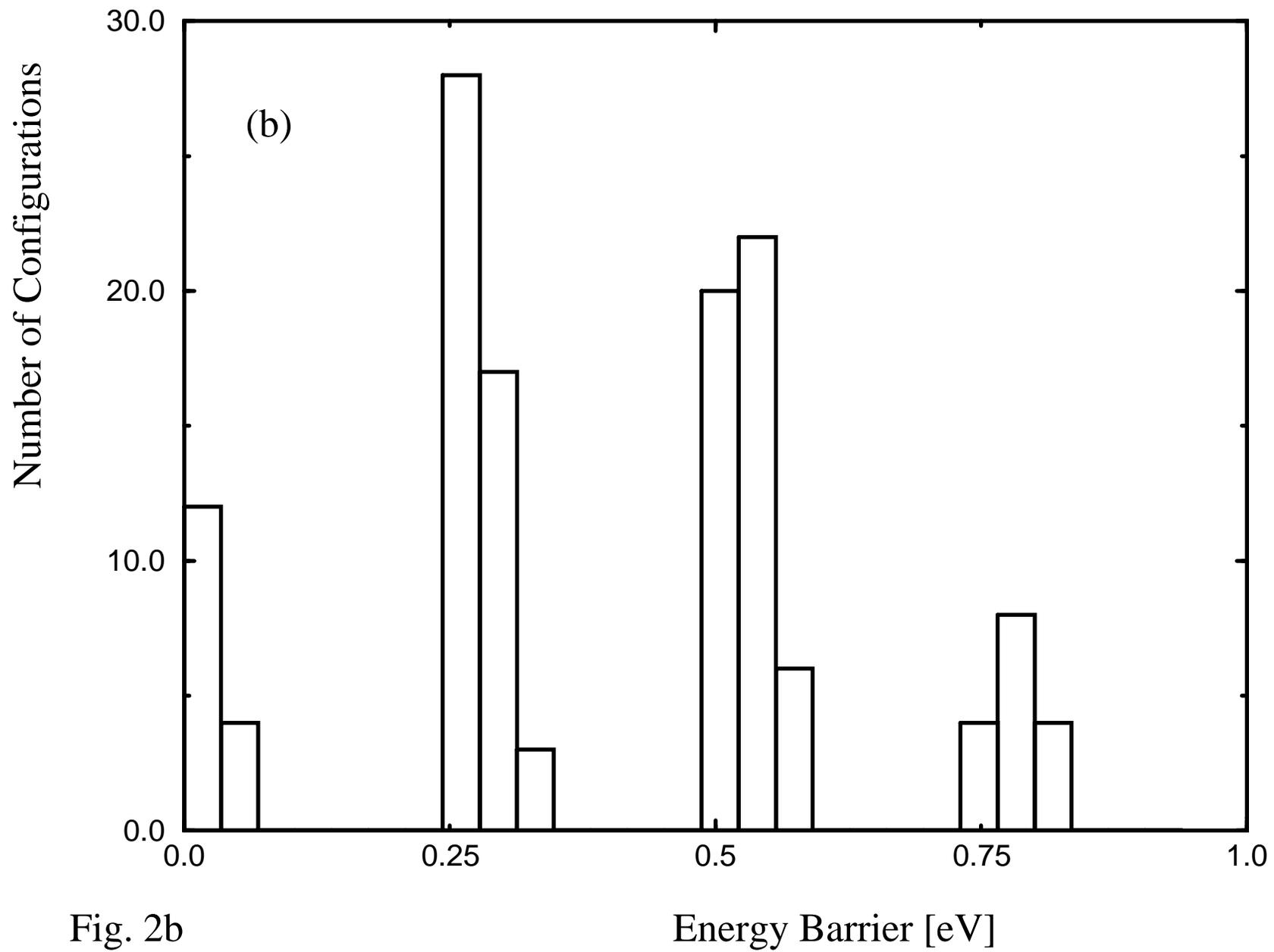
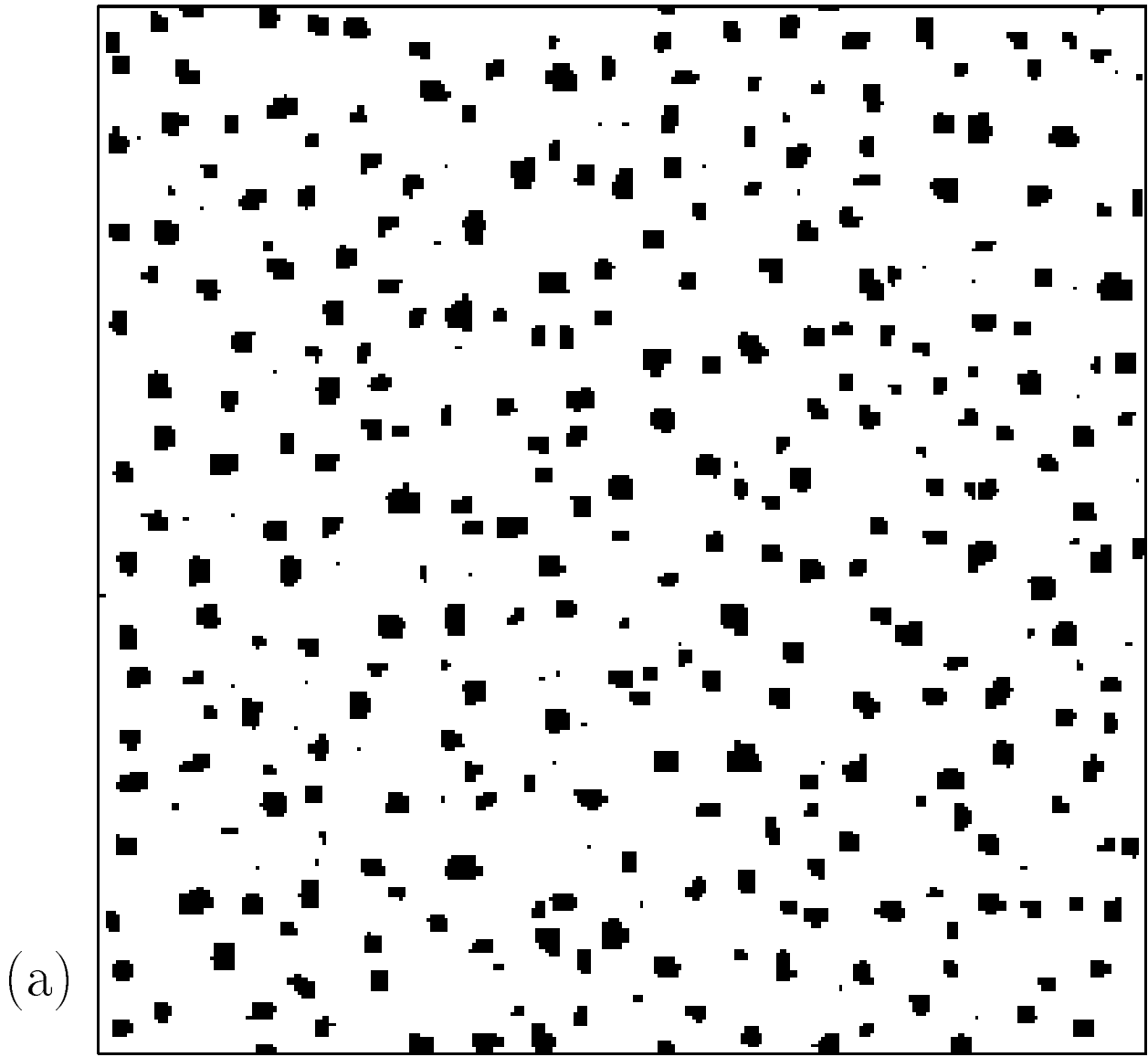
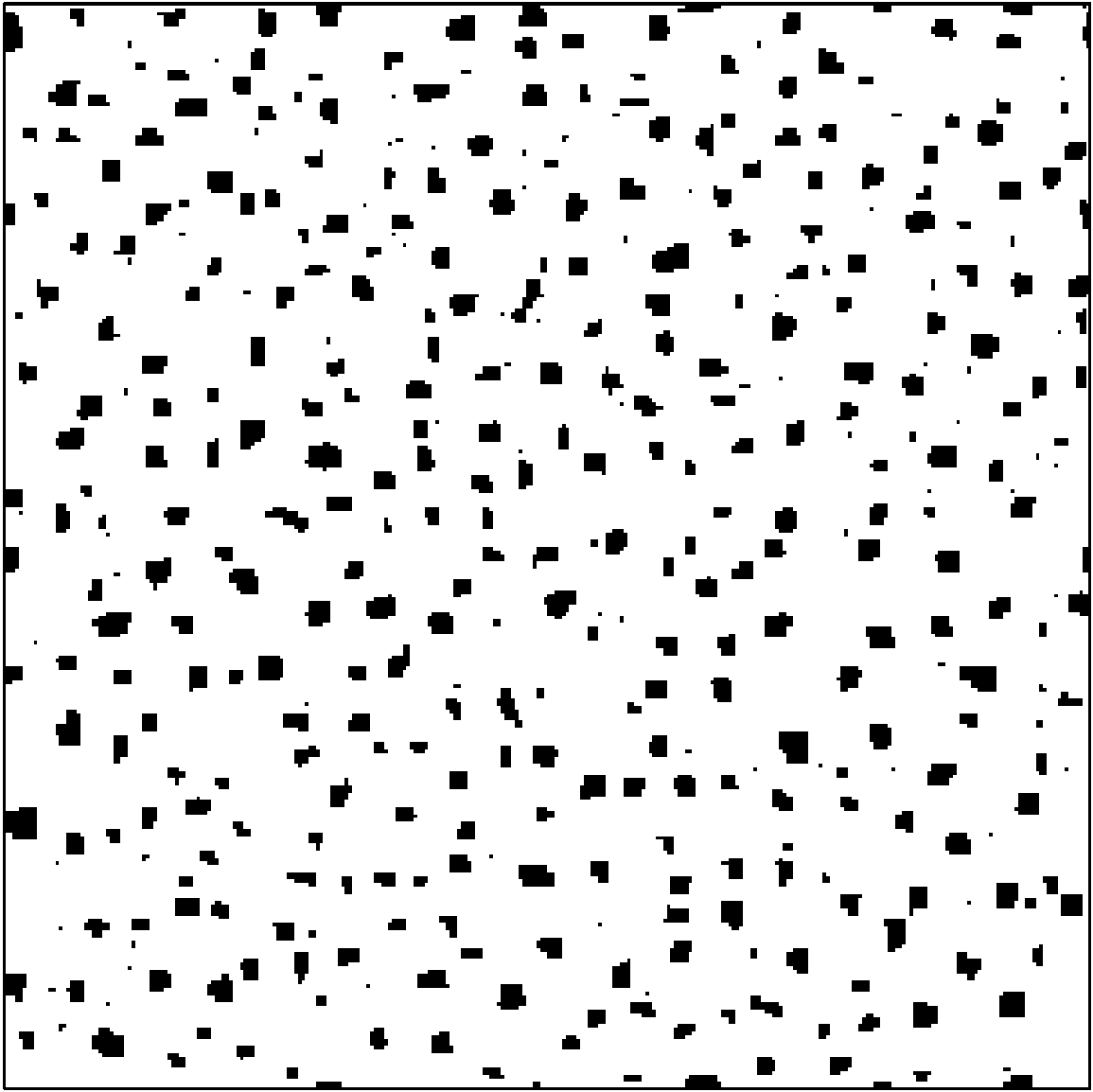
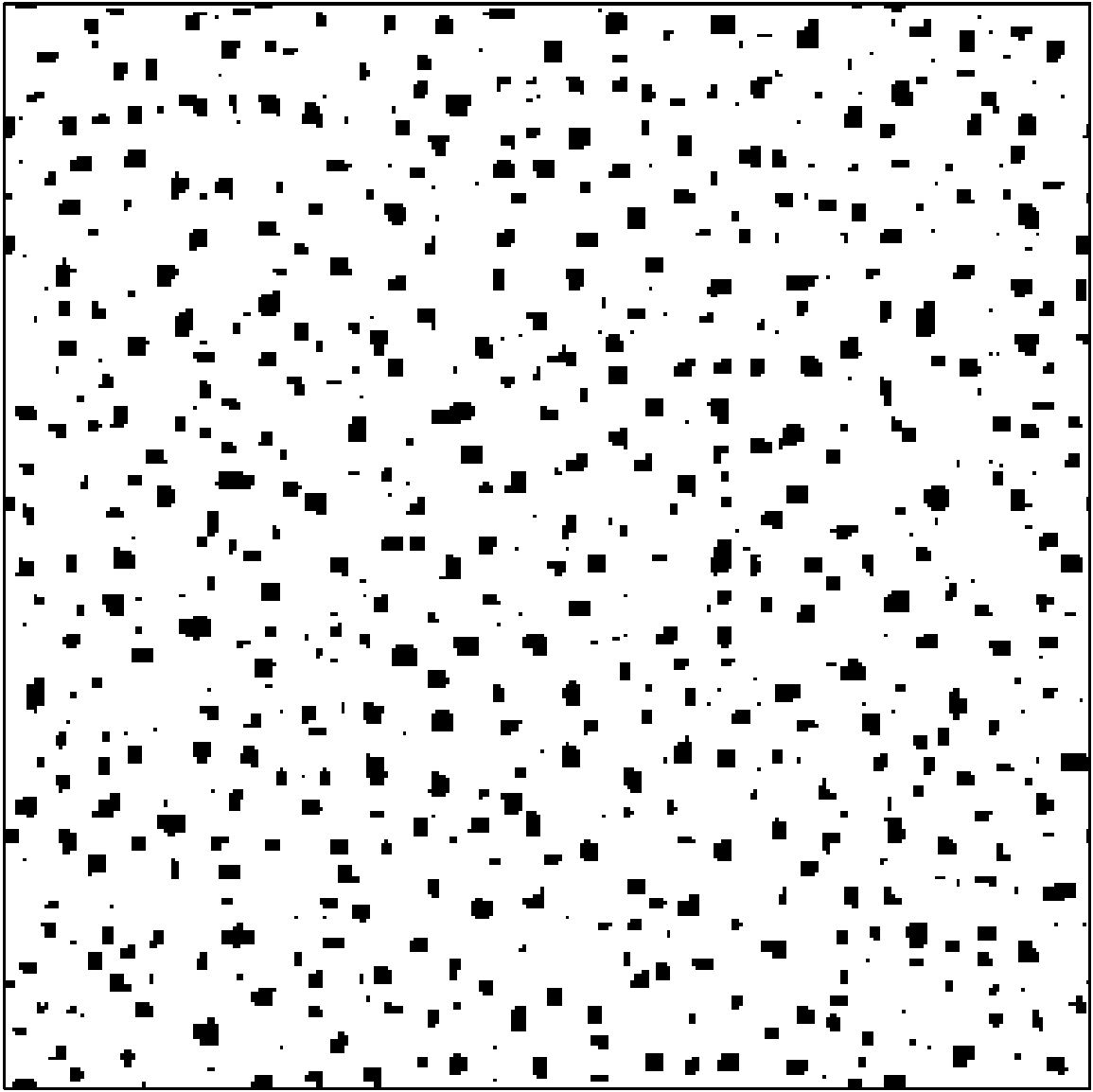
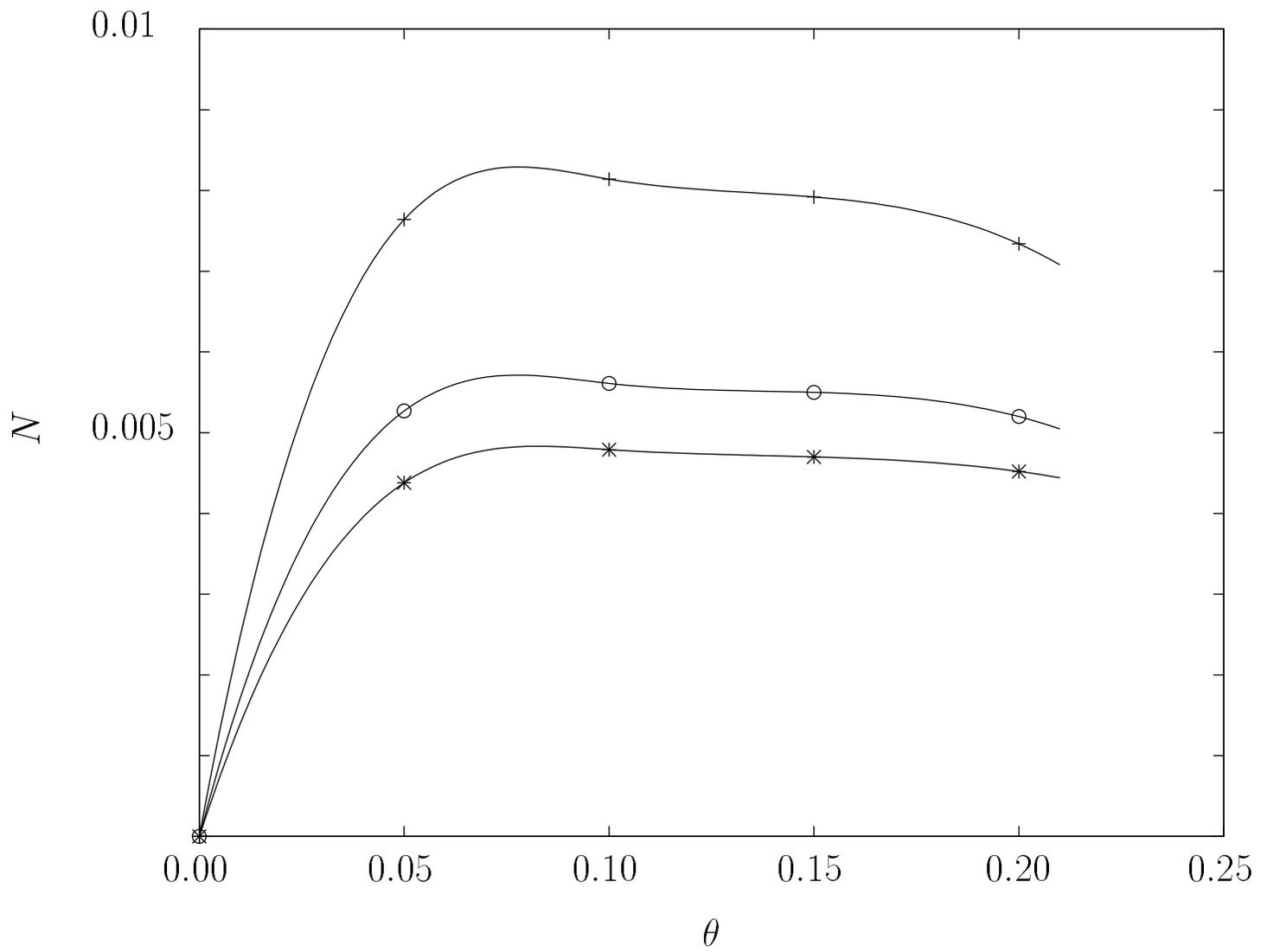
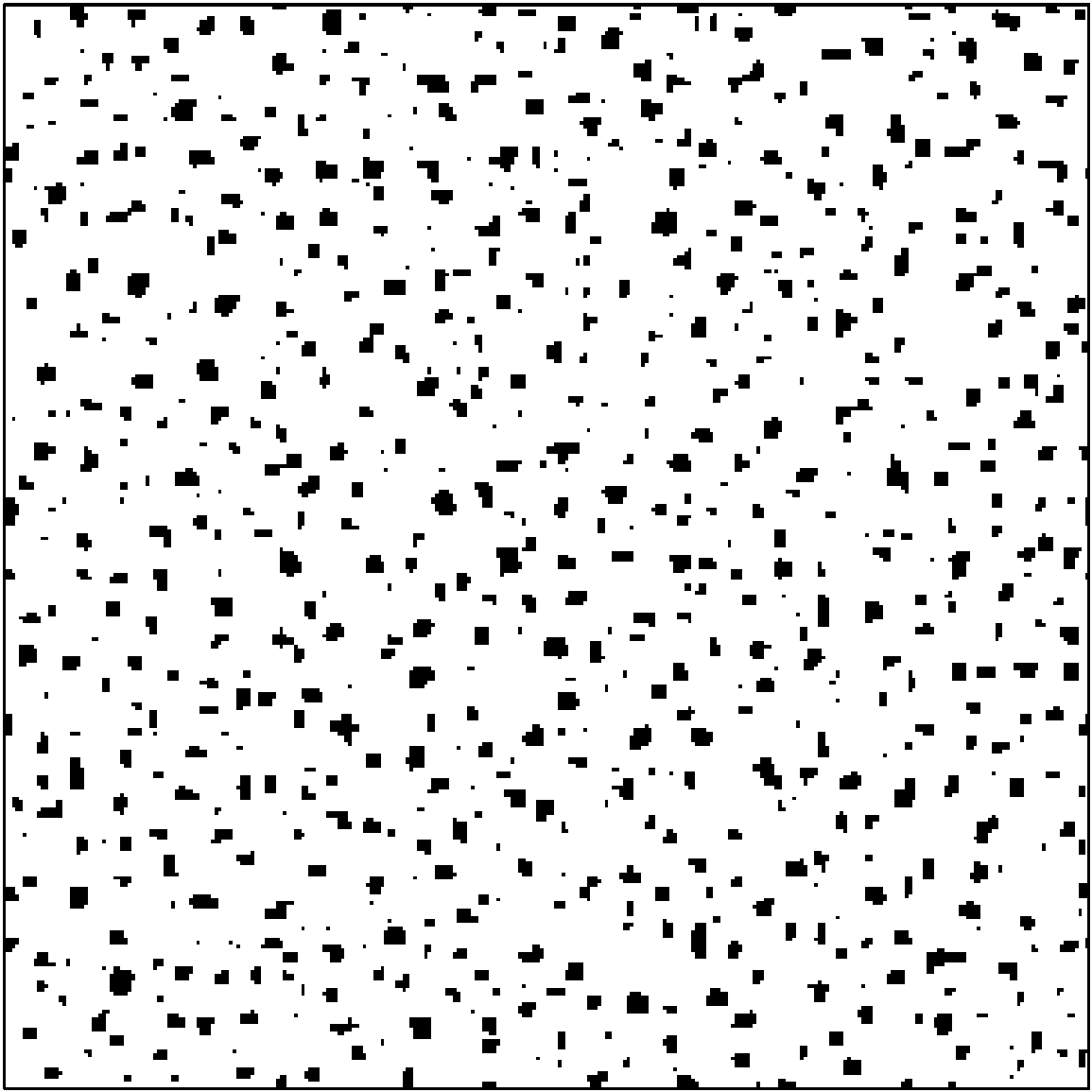
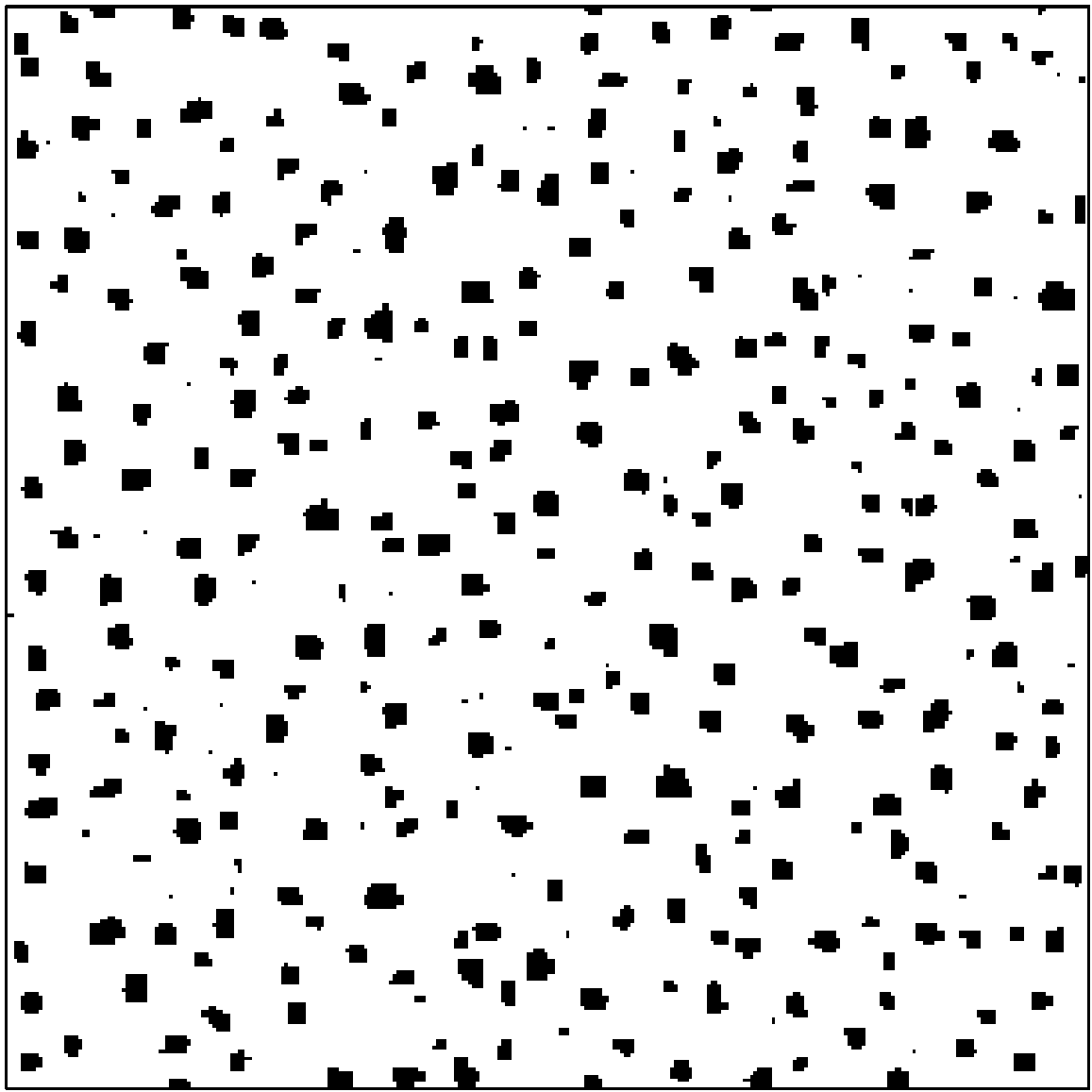
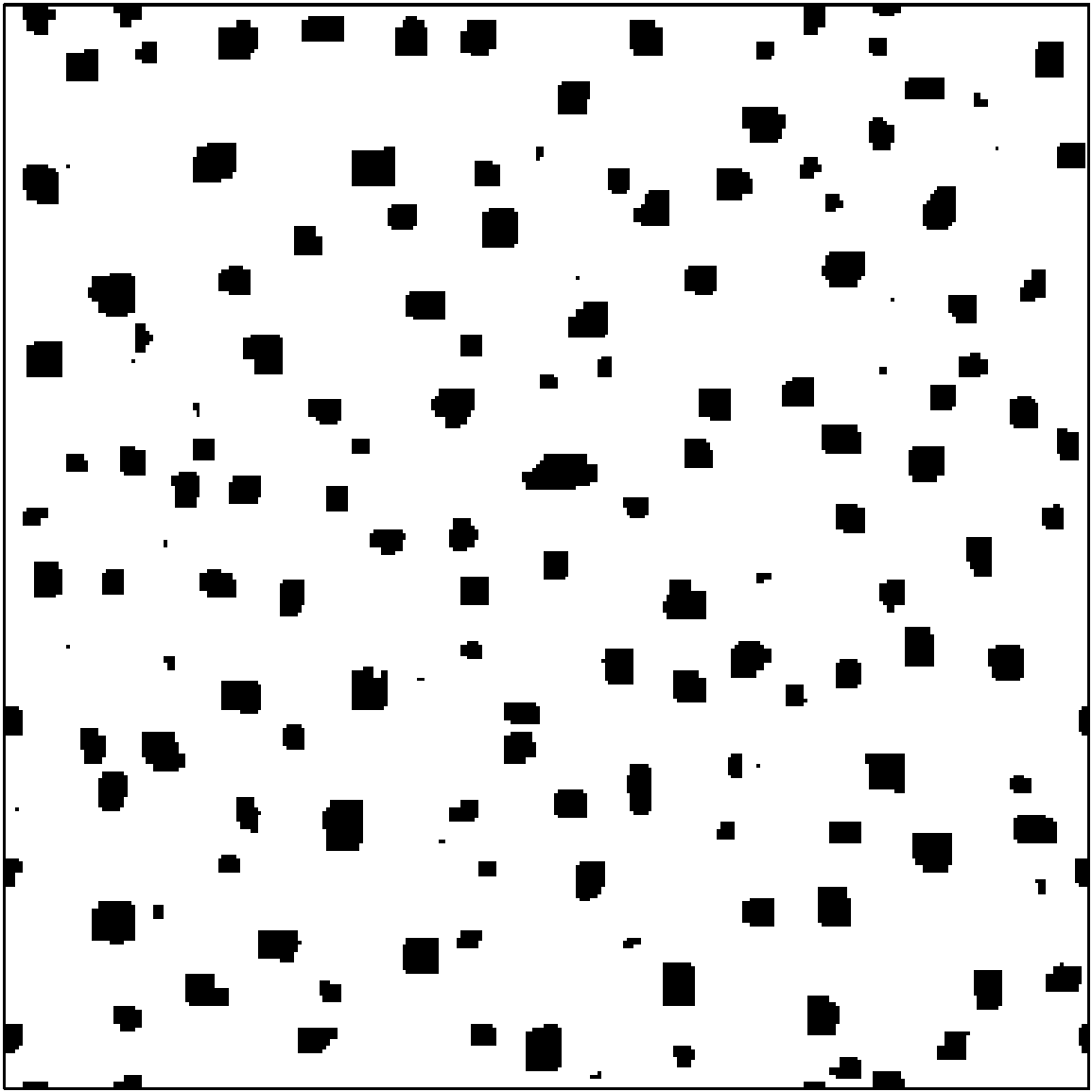
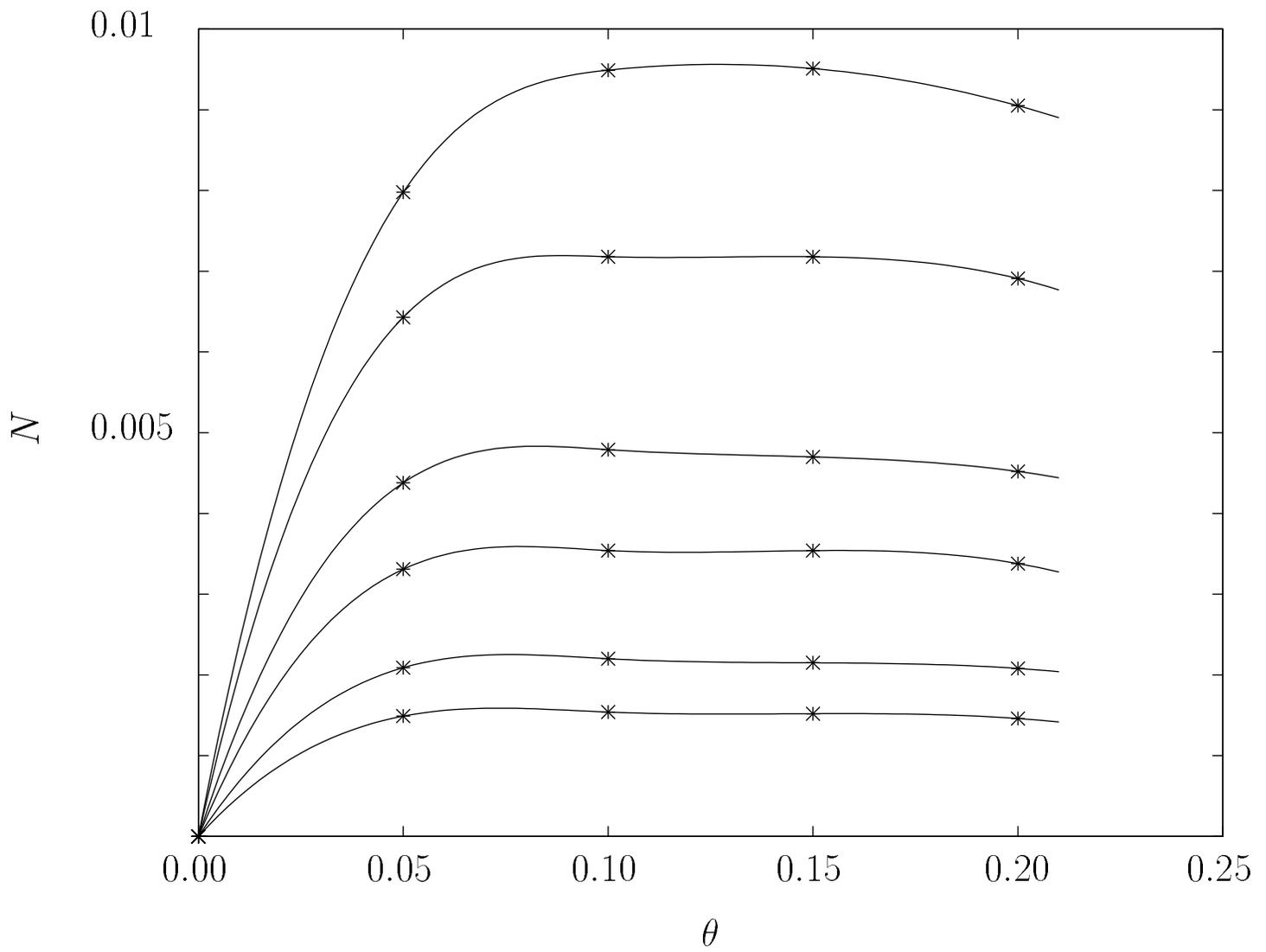
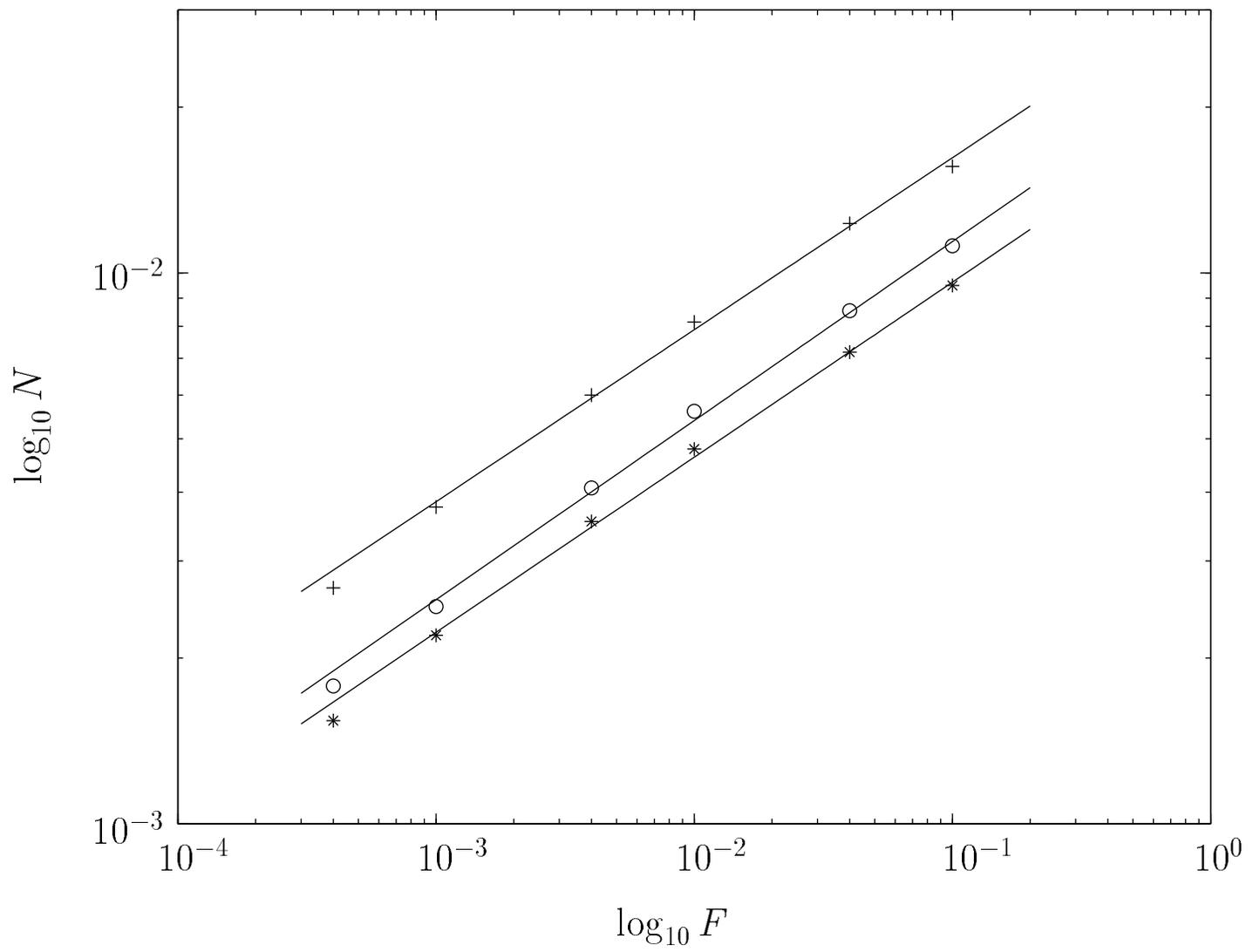
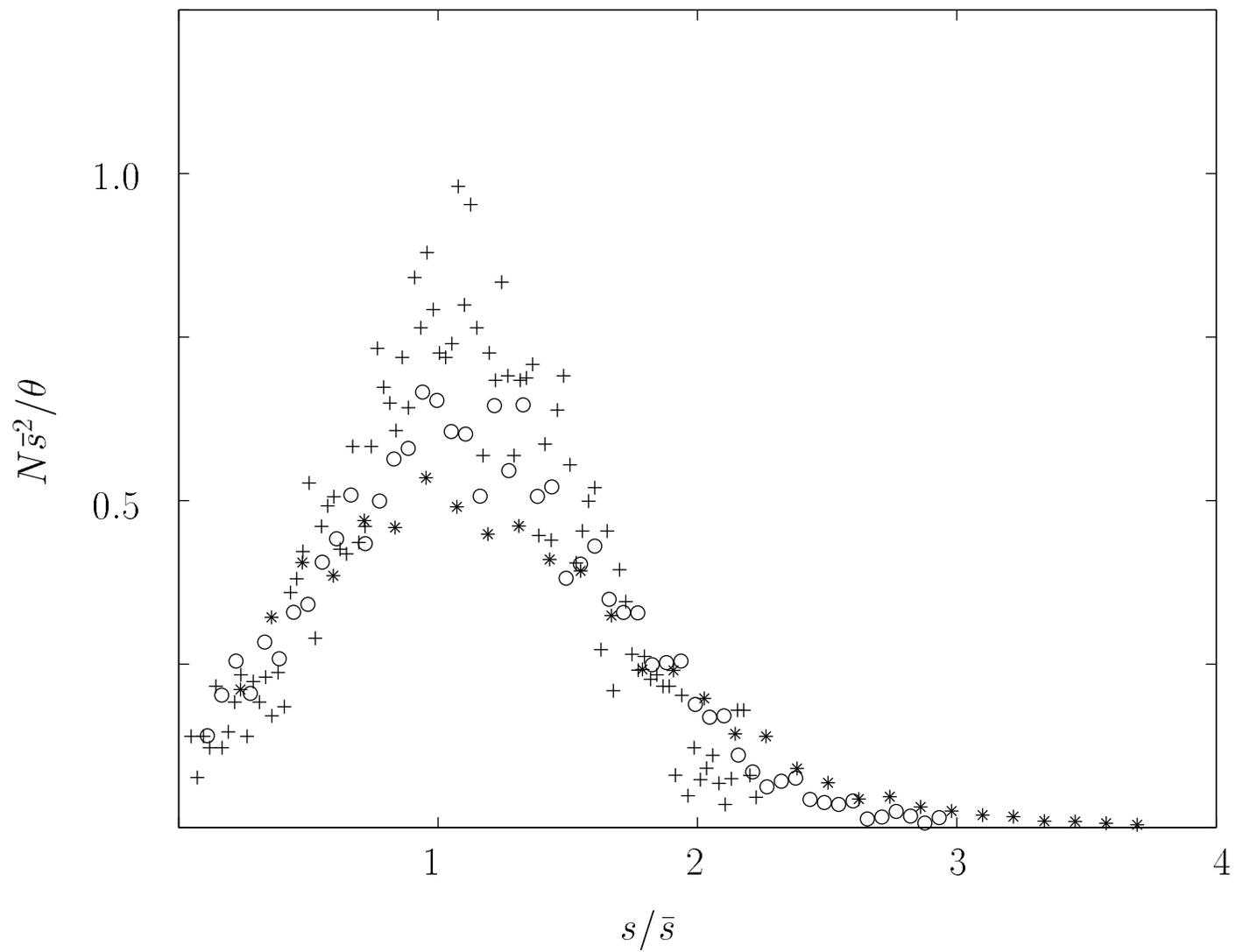
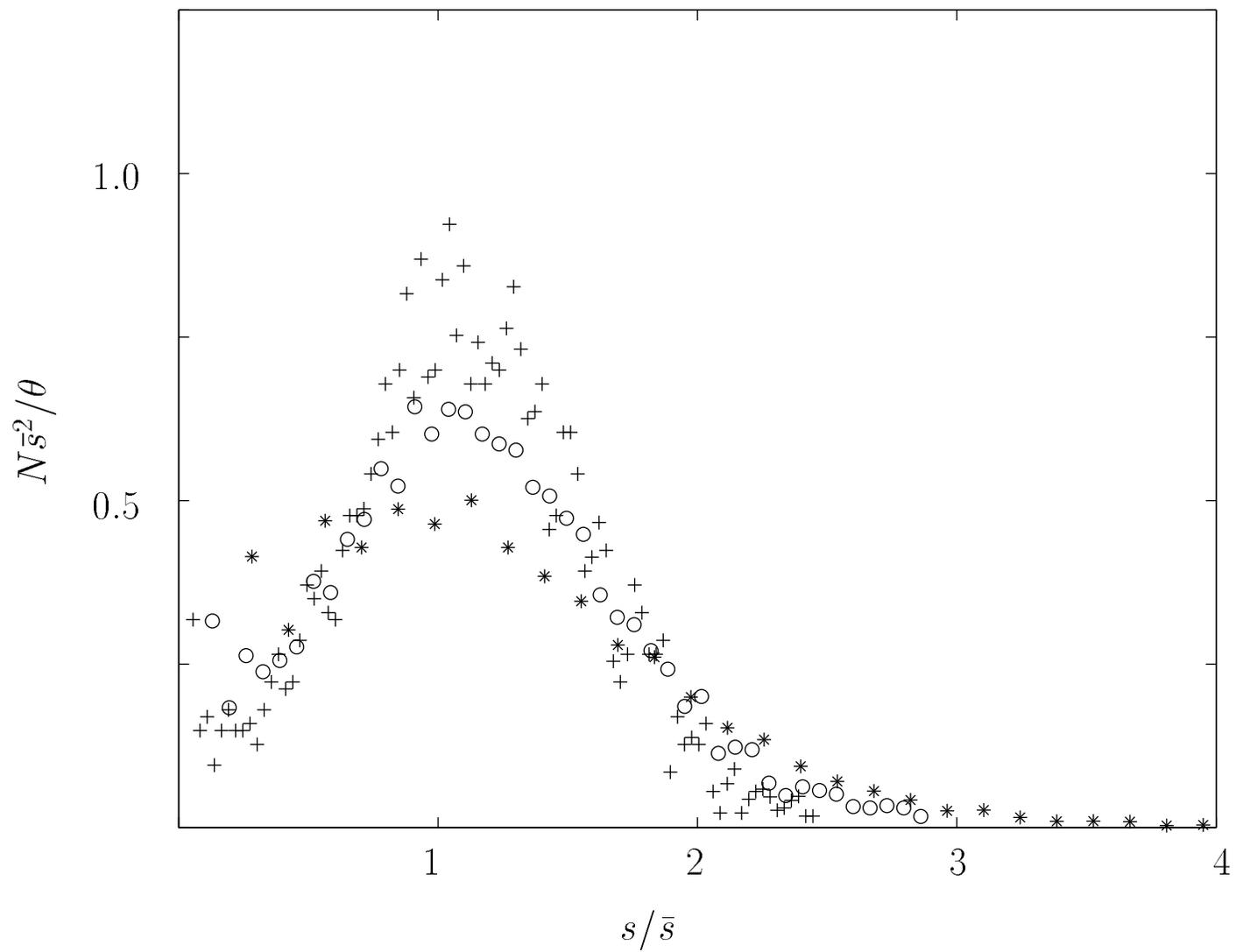
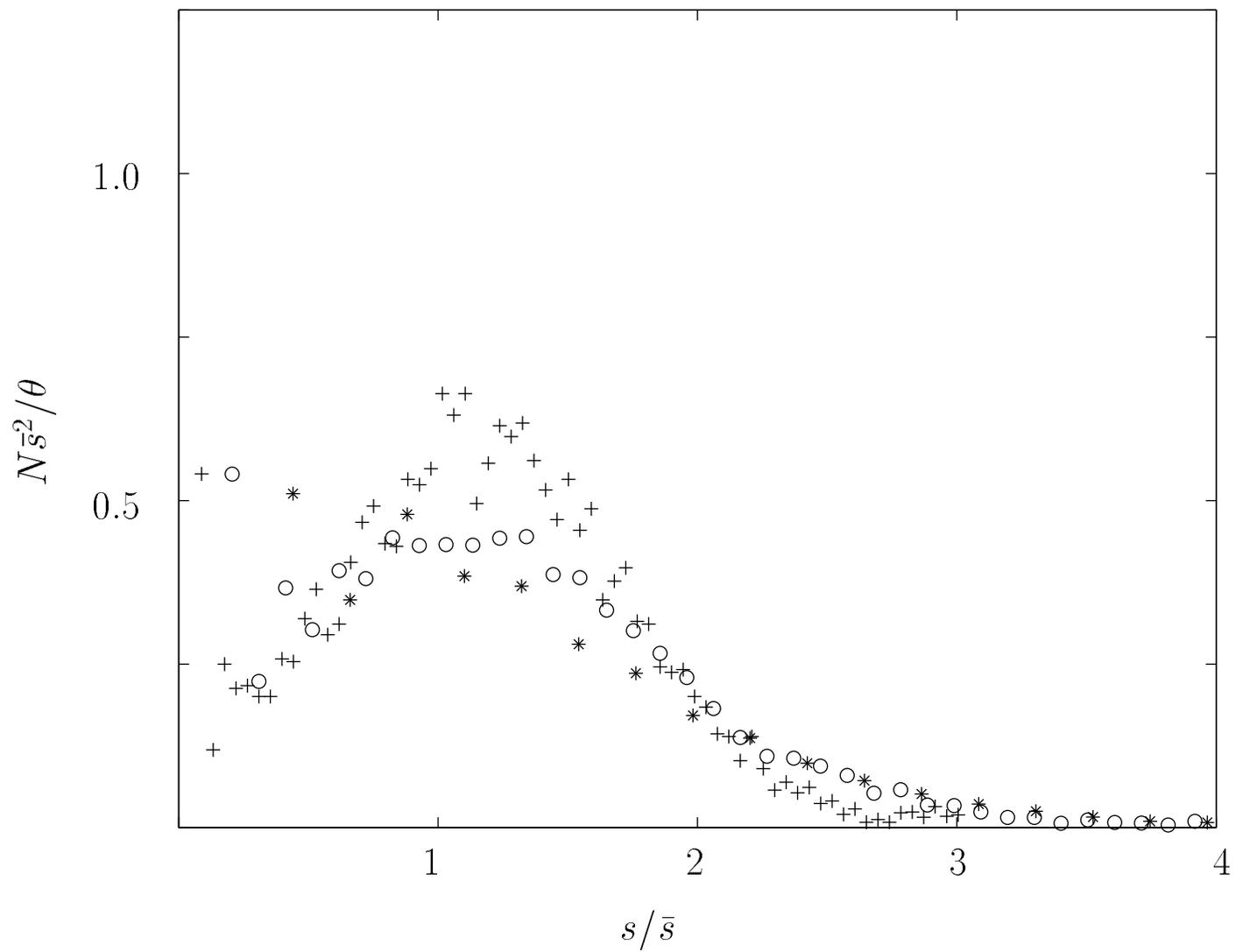
\end{document}